\documentclass[fleqn,usenatbib,useAMS]{mnras}
\usepackage{graphicx}
\usepackage{amsmath}
\usepackage{amsfonts}
\usepackage{float}
\usepackage{bm}
\usepackage[perpage,symbol*]{footmisc}
\usepackage[T1]{fontenc}
\usepackage{ae,aecompl}
\usepackage{array}

\newcommand{\appropto}{\mathrel{\vcenter{
  \offinterlineskip\halign{\hfil$##$\cr 
    \propto\cr\noalign{\kern2pt}\sim\cr\noalign{\kern-2pt}}}}}

\newcommand{\ssim}{\,{\sim}\,} 

\makeatletter

\newcommand{\Rmnum}[1]{\expandafter\@slowromancap\romannumeral #1@}
\makeatother

\title[The 3D Local Group Timing Argument In $\Lambda$CDM]{Dynamical History Of The Local Group In $\Lambda$CDM \Rmnum{2} $-$ Including External Perturbers In 3D}
\author[Indranil Banik \& Hongsheng Zhao]{Indranil Banik$^{1}$\thanks{Email: \href{mailto:ib45@st-andrews.ac.uk}{ib45@st-andrews.ac.uk} (Indranil Banik)\newline $~~~~~~~~~~~~~~$ \href{mailto:hz4@st-andrews.ac.uk}{hz4@st-andrews.ac.uk} (Hongsheng Zhao)}, Hongsheng Zhao$^{1}$\\
$^{1}$Scottish Universities Physics Alliance, University of St Andrews, North Haugh, St Andrews, Fife, KY16 9SS, UK}

\pubyear{2017}
\begin{document}
\label{firstpage}
\pagerange{\pageref{firstpage}--\pageref{lastpage}}

\maketitle

\begin{abstract}



We attempt to fit the observed radial velocities (RVs) of $\ssim$30 Local Group (LG) galaxies using a 3D dynamical model of it and its immediate environment within the context of the standard cosmological paradigm, $\Lambda$CDM. This extends and confirms the basic results of our previous axisymmetric investigation of the LG (MNRAS, 459, 2237). We find that there remains a tendency for observed RVs to exceed those predicted by our best-fitting model. The typical mismatch is slightly higher than in our 2D model, with a root mean square value of $\ssim 50$ km/s. \emph{Our main finding is that including the 3D distribution of massive perturbing dark matter halos is unlikely to help greatly with the high velocity galaxy problem.} Nonetheless, the 2D and 3D results differ in several other ways such as which galaxies' RVs are most problematic and the preferred values of parameters common to both models.

The anomalously high RVs of several LG dwarfs may be better explained if the Milky Way (MW) and Andromeda (M31) were once moving much faster than in our models. This would allow LG dwarfs to gain very high RVs via gravitational slingshot encounters with a massive fast-moving galaxy. Such a scenario is possible in some modified gravity theories, especially those which require the MW and M31 to have previously undergone a close flyby. In a $\Lambda$CDM context, however, this scenario is not feasible as the resulting dynamical friction would cause a rapid merger.

\end{abstract}


\section{Introduction}
\label{Introduction}








The dynamics of the Local Group (LG) of galaxies provided an early indication that our current understanding of physics is insufficient to explain the dynamics of astrophysical systems. Although the Universe must have started off expanding, the Andromeda (M31) and Milky Way (MW) galaxies are currently approaching each other at  $\sim$110 km/s \citep{Slipher_1912, Schmidt_1958}. Their initial recession could not have been turned around in the $\ssim 14$ Gyr \citep{Planck_2015} since the Big Bang if the luminous masses of these galaxies attract each other according to the inverse square law of Newtonian gravity \citep{Kahn_Woltjer_1959}.

The most commonly accepted solution is that most galaxies $-$ including the MW and M31 $-$ are surrounded by large amounts of dark matter \citep[e.g.][]{Ostriker_Peebles_1973}. For a while, it was thought that this could be non-luminous conventional matter such as stellar remnants \citep[e.g.][and references therein]{Carr_1994}. However, gravitational microlensing searches for such massive compact halo objects indicated that there was not enough mass in them \citep{MACHO_2000, EROS_2007}. Thus, the required dark matter is thought to consist of an undiscovered stable particle, or at least one with a decay time longer than the age of the Universe \citep[e.g.][and references therein]{Steigman_1985}. Multi-decade searches for this particle have now ruled out a substantial part of the parameter space which was thought to be feasible before the searches started \citep[e.g.][]{Ackerman_2015, LUX_2016, PandaX_2016}.

Knowing only the separation and relative velocity of two galaxies, it would be very difficult to rule out this scenario. These two pieces of information are sufficient to constrain the relevant model parameters: the initial MW$-$M31 co-moving separation and their combined mass, some of which would lie beyond their virial radii \citep{Fattahi_2016}.\footnote{We assume throughout that the masses of all galaxies do not change over time.} Fortunately, much additional information has recently become available in the form of positions and velocities of many other LG galaxies \citep[e.g.][and references therein]{McConnachie_2012}.

The velocity field traced out by these galaxies should be understandable using the same MW and M31 total mass as is required to explain their present relative motion. An early attempt at such an analysis was made by \citet{Sandage_1986}. There was some difficulty in matching all the data then available.

A more recent analysis based primarily on the catalogue of \citet{McConnachie_2012} also treated the gravitational field of the LG as spherically symmetric \citep{Jorge_2014}. Later, an adjustment was made for the effect of the Large Magellanic Cloud (LMC) on the MW, and thus on the observed velocities of all galaxies in the LG \citep{Jorge_2016}.

The MW and M31 can only be treated as a single point mass if their mutual separation $d$ \citep[${783 \pm 25}$ kpc,][]{McConnachie_2012} is much less than the distance to the galaxy one is interested in. However, the LG only extends out to $\ssim 3$ Mpc, making this assumption not very accurate within it. Further away, other massive objects besides the MW and M31 must also be considered.

\citet{Jorge_2014} considered the effects that the MW-M31 quadrupole might have, to get the lowest order correction to the spherical symmetry assumption in their work (see their Section 2.4). However, at distances from the LG barycentre of only $\ssim 2 \times$ the MW-M31 separation, it is likely that higher order terms would also be important. Moreover, the quadrupole term was not rigorously included in their final analysis.

Thus, we constructed an axisymmetric model of the LG in $\Lambda$CDM. We consider this reasonable because the low proper motion of M31 suggests an almost radial MW$-$M31 orbit \citep{M31_motion}. A major nearby perturber to the LG is the Centaurus A group of galaxies \citep{Harris_2010}. This lies very close to the MW$-$M31 line, allowing us to incorporate it into our model. Therefore, our simulation included 3 massive objects, with LG dwarfs treated as test particles.

We previously published results based on this 2D model \citep{Banik_Zhao_2016} and review it here (Section \ref{2D_review}). Motivated by a poor match between the model and observations, we consider a 3D model of the LG (Section \ref{3D_method}). The results obtained using it are described in Section \ref{3D_results}. We go on to discuss how they compare with those obtained using our 2D model (Section \ref{Discussion}). Here, we also consider a few factors beyond those directly included in our models. Section \ref{Great_Attractor} is devoted to the effects of the Great Attractor and the Virgo Cluster on the LG. In Section \ref{MOND}, we discuss the possible effects of a departure from the Newtonian gravity law assumed elsewhere in this work. Our conclusions are provided in Section \ref{Conclusions}.

\section{Review of 2D axisymmetric model}
\label{2D_review}

\subsection{Governing Equations}

We begin by reviewing our axisymmetric dynamical model of the LG \citep{Banik_Zhao_2016}, which in turn follows on from the earlier spherically symmetric analysis of \citet{Jorge_2014}. Our simulations start at a redshift of 9, when the expansion of the Universe was nearly homogeneous \citep{Planck_2015}. Thus, we assume that everything was following a smooth Hubble flow at that time. This means that the velocity $\bm{v}$ of each simulated particle would depend on its position $\bm{r}$ according to
\begin{eqnarray}
	\bm{v_{_i}} ~=~ H_{_i} \bm{r}_{_i}
	\label{Initial_conditions}
\end{eqnarray}

For any quantity $k$, we use $k_{_i}$ to denote its value at the time when our simulations are started and $\dot{k}$ to denote its time derivative. The expansion rate of the Universe is quantified by the Hubble parameter $H \equiv \frac{\dot{a}}{a}$, where $a$ is the cosmic scale-factor. At the present time, $H = H_{_0}$ and $a =1$. In a flat Universe containing only matter and dark energy, their values at other times are given implicitly by
\begin{eqnarray}
	H\left( t \right) ~=~ H_{_0} \sqrt{\frac{{\Omega }_{m ,0}}{a^3\left( t \right)} ~+~ {\Omega }_{\Lambda ,0}}
	\label{H}
\end{eqnarray}

We use a standard flat ($\Omega_{m,0} + \Omega_{\Lambda, 0} \equiv 1$) dark energy-dominated cosmology whose parameters are given at the bottom of Table \ref{Best_fit_parameters}. Defining time $t$ to start at the Big Bang (${a = 0}$) and imposing the current expansion rate of the Universe as a boundary condition, we get that
\begin{eqnarray}
	a(t) &=& {{\left( \frac{{{\Omega }_{m,0}}}{{\Omega }_{\Lambda ,0}} \right)}^{\frac{1}{3}}}{{\sinh }^{\frac{2}{3}}}\left( \frac{3}{2}\sqrt{{{\Omega }_{\Lambda ,0}}}~{{H}_{_0}}t \right)
\label{Expansion_history}
\end{eqnarray}

LG dwarf galaxies are represented as test particles affected by the expansion of the Universe and by three massive particles $-$ the MW, M31 and Centaurus A. The dynamics of test particles in such situations can be understood using General Relativity \citep[][Section 2.1]{Banik_Zhao_2016}.

We constrain the massive particles to move along a line, making our model axisymmetric. Starting with a plane-polar grid of initial positions, we advance the trajectories of a large number of test particles using the equation of motion
\begin{eqnarray}
	{\overset{..}{\bm r} } = {\frac{\overset{..}{a}}{a}}  {\bm r} ~- \sum_{\begin{array}{c} \text{j = MW,}\\ \text{M31, Cen A}\end{array}}{ \frac{G M_j \left( \bm r - \bm r_{_j} \right)} {{\left( |\bm r - \bm r_{_j} |^2 + {r_{_{S,j}}}^2 \right)}^{\frac{1}{2}} |{\bm r}- \bm r_{_j} |^2}}
	\label{Equation_of_motion_2D}
\end{eqnarray}

\begin{figure}
	\centering 
		\includegraphics [width = 8.5cm] {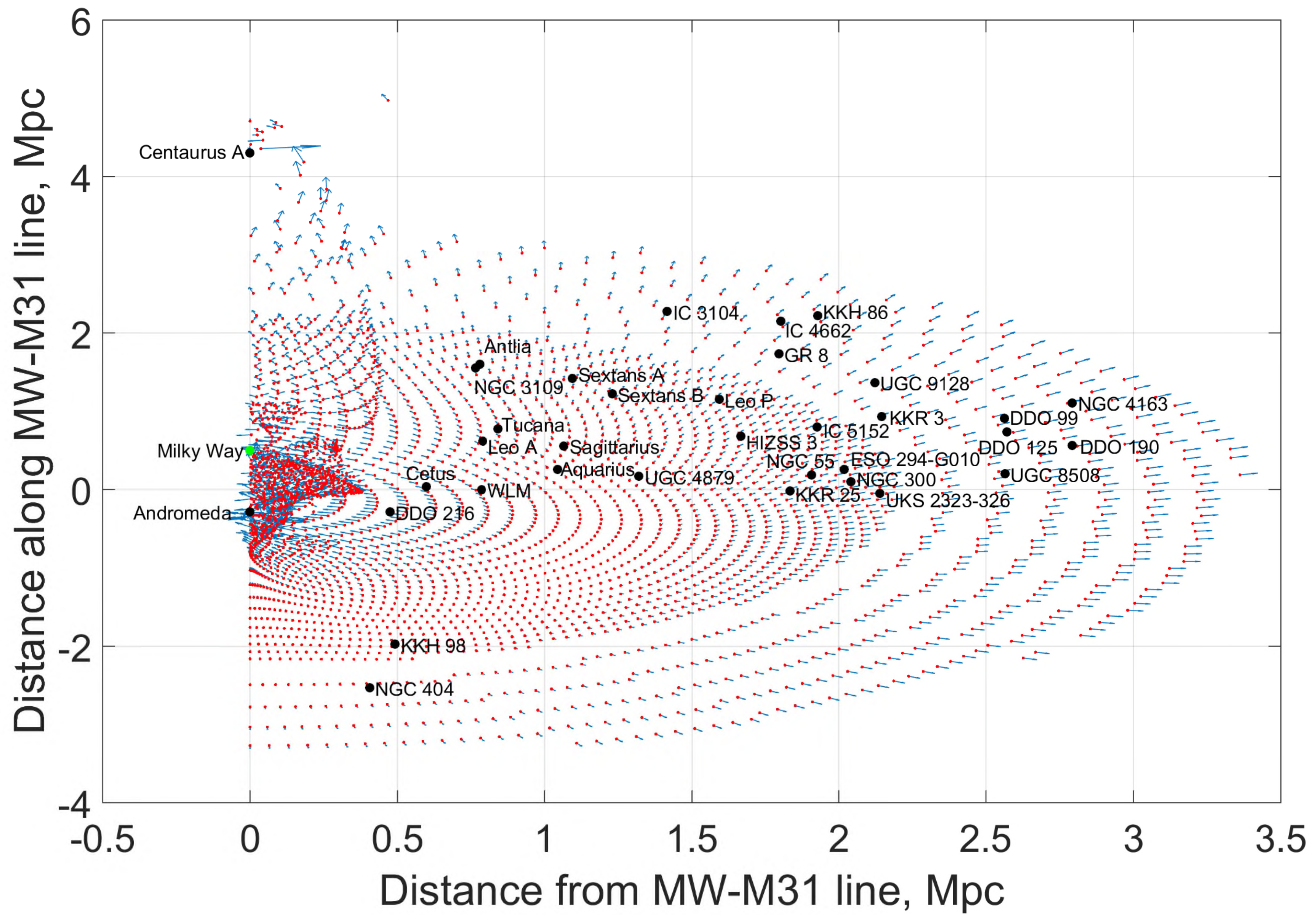}
		\includegraphics [width = 8.5cm] {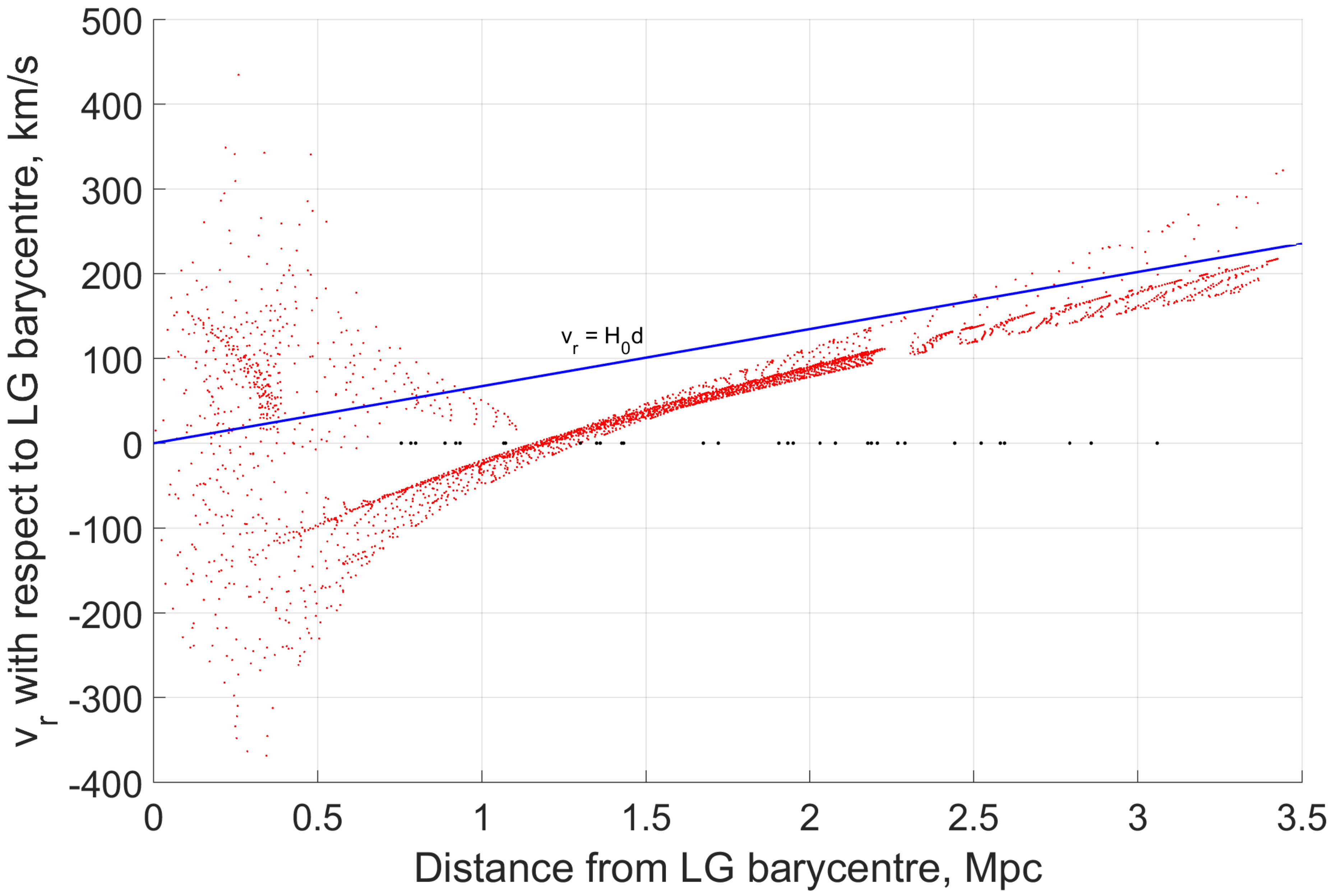}
	\caption{\emph{Top:} Local Group velocity field for our best-fitting axisymmetric simulation (parameters given in Table \ref{Best_fit_parameters}). The radial resolution was degraded beyond a distance of 2.3 Mpc as the velocity field is generally quite smooth there. Locations of indicated galaxies are shown relative to the MW$-$M31 line. \emph{Bottom:} Radial velocities of test particles with respect to the LG barycentre. Black dots on the $x$-axis show the distances of target galaxies from there. Without proper motions, observations can't be put on such a Hubble diagram because the MW is not at the LG barycentre.}
	\label{Velocity_field_2D_LMC}
\end{figure}

$r_{_S}$ is chosen so that the force at $r \ll r_{_S}$ leads to the correct flatline level of rotation curve for each galaxy, i.e. $r_{_S} = \frac{GM}{{v_{_f}}^2}$ where $M$ is the mass of the relevant galaxy, whose rotation curve flatlines at the level $v_{_f}$. For the MW, we take $v_{_f} = 180$ km/s \citep{Kafle_2012} while for Andromeda, we use $v_{_f} = 225$ km/s \citep{Carignan_2006}. Fixing $v_{_f}$ meant that we had to adjust $r_{_S}$ for the MW and M31 depending on their respective masses. Because separations between the massive galaxies are always quite large, for simplicity we use a pure inverse square law for the forces between them (i.e. $r_{_S}$ is set to 0 when calculating forces between the three massive particles). For test particles that get within $\ssim 15$ kpc of a massive galaxy\footnote{31 kpc for Centaurus A}, we simply terminate the trajectory. It is likely that any real LG dwarf in this situation would be severely disrupted. Moreover, our analysis is concerned with LG dwarfs much further from any of the three massive galaxies in our model (Figure \ref{Velocity_field_2D_LMC}).

Our algorithm advances trajectories using a fourth order Runge-Kutta method based on an adaptive but quantised timestep, ensuring that the positions of the massive particles are available when needed. The timestep is adapted based on the distances between the object being advanced and the massive galaxies which influence its motion.

Our model is designed to accurately match the presently observed positions of all galaxies within the LG as well as Centaurus A. Thus, we used a 2D Newton-Raphson algorithm to vary the initial relative positions of the MW, M31 and Cen A along a line in order to match their presently observed configuration to within an accuracy of $\ssim 10^{-4}$. Solutions involving collisions between any of these galaxies were of course discarded. We were able to obtain a valid solution in all cases, though this was much easier if the algorithm was slightly under-relaxed for better stability.

Once the trajectories of the massive objects were known, we used Equation \ref{Equation_of_motion_2D} to advance test particle trajectories. The resulting velocity field in one of our simulations is shown in Figure \ref{Velocity_field_2D_LMC}.

To get a test particle landing very close to the observed position of each LG dwarf galaxy, we could simply use a very dense grid of initial positions for our test particles. However, this would be very computationally intensive, especially if extended to 3 dimensions. Thus, we used a modest resolution grid and found which test particle landed closest to each target galaxy. We then used a 2D Newton-Raphson algorithm to vary the initial position of this test particle, targeting the presently observed position of the corresponding galaxy. Because varying the trajectory of a test particle does not alter the gravitational field in the LG, we were able to improve the accuracy slightly, to $\ssim 10^{-5}$. 

The present velocity of the test particle on this trajectory is our model prediction for the velocity of the target galaxy it represents. We subtract the simulated velocity of the MW and then project the relative velocity onto the direction towards the target to get its model-predicted Galactocentric Radial Velocity (GRV).

To convert observed heliocentric radial velocities (HRVs) into Galactocentric ones, we need an independent constraint on $\bm v_\odot$, the motion of the Sun within the MW. Part of the challenge is determining $v_{c, \odot}$, the speed of a test particle on a circular orbit around the MW at the position of the Sun. This is called the Local Standard of Rest (LSR). The other part is the non-circular component of $\bm v_\odot$, which consists of $U_\odot$ towards the Galactic Centre, $V_\odot$ in the direction of rotation and $W_\odot$ towards the North Galactic Pole. Our adopted values for these parameters are given in Table \ref{Best_fit_parameters}, the caption of which contains the relevant references. Given this information, we can determine actual GRVs using the relation
\begin{eqnarray}
	GRV_{obs} ~=~ HRV_{obs} ~+~ \bm v_\odot \cdot \widehat{\bm d}_{_{MW}}
	\label{GRV_obs}
\end{eqnarray}

We use $\widehat{\bm d}_{_{MW}}$ for the direction from the MW towards a target galaxy. The term $\left( \bm v_\odot \cdot \widehat{\bm d}_{_{MW}} \right)$ represents a correction for Solar motion within the MW. Because $v_{c, \odot}$ is a model parameter, this correction is slightly model-dependent.


At large distances from the MW, it and M31 may be considered as a single point mass $M$. However, even in this case, the MW mass fraction $q_{_{MW}} \equiv \frac{M_{_{MW}}}{M}$ has a substantial effect on GRVs. This arises because smaller values of $q_{_{MW}}$ imply that the MW is moving faster with respect to the LG barycentre. As a result, even a spherically symmetric model of the LG can be used to place meaningful constraints on $q_{_{MW}}$, as was recently done by \citet{Jorge_2014}.

We have implicitly assumed that the motion of the Sun with respect to the disk of the MW is the same as its motion with respect to the MW barycentre, the important quantity for our timing argument analysis. This assumption may be invalidated if the MW has massive satellite galaxies. In fact, this does seem to be the case, especially when considering the LMC \citep{Jorge_2016}.


In our models, the LMC is not treated as another particle but as part of the MW. Thus, its simulated mass includes that of the LMC. Effectively, our model uses one particle to represent the MW system ($\equiv$ MW $+$ satellite). This assumes that the LMC is bound to the MW, whose disk must then be moving with respect to the barycentre of the MW system due to the recoil induced by the LMC. Because it is very nearby compared to other LG galaxies of interest, we neglect the fact that observations made from near the Sun are no longer made at the barycentre of the MW system.\footnote{We take the Sun to be at the barycentre of the MW system. However, our 3D model (Section \ref{3D_method}) accounts for the LMC directly and treats the Sun as being 8 kpc from the centre of the MW disk.}

Although such position effects are negligible, it is important to consider the motion of the LMC. Thus, we determined $\bm v_{_{LMC}}$, its space velocity with respect to the MW. This requires knowledge of its HRV \citep{McConnachie_2012} and its proper motion \citep{Kallivayalil_2013} multiplied by its distance \citep{Pietrzynski_2013}. This information was used to obtain a revised estimate for the motion of the Sun with respect to the MW system
\begin{eqnarray}
	\bm v_\odot &\to& \bm v_\odot ~-~ q_{_{LMC}} \bm v_{_{LMC}} \\
	q_{_{LMC}} &\equiv& \frac{M_{LMC}}{M_{MW}} ~~\text{ (MW includes LMC) }
	\label{LMC_adjustment}
\end{eqnarray}

Although M31 may have massive satellites too, we do not consider them because they do not affect our analysis to the same extent. A massive satellite of M31 can create a mismatch between the present GRV of the M31 disk and that of the M31 system. Our model would account for the mass of the M31 satellite as part of the M31 mass itself. The main effect on galaxies other than M31 would be a shift in the barycentre of the M31 system by a few kpc. This should have only a very small effect on the dynamics of other galaxies due to their large distances from M31 (top panel of Figure 1). Consequently, only the GRV of M31 could be noticeably affected in this scenario. As M31 is only one of our 34 target galaxies, our analysis should not change much overall. This is not true if the MW had a massive satellite as that would affect our velocity relative to everything else.

\subsection{Statistical analysis}
\label{2D_statistical_analysis}

We used our axisymmetric model to predict GRVs of target galaxies where it was possible to obtain a unique prediction. This is not always the case, as is clear from Figure \ref{Velocity_field_2D_LMC}. In the region between the MW and M31, intersecting trajectories are apparent. This means that there is more than one possible velocity at the same position, even in the same model. Thus, we do not have any targets within this region. Based on this consideration, we made some adjustments to the catalogue of galaxies used by \citet{Jorge_2014} for our analysis in \citet{Banik_Zhao_2016}. In particular, we excluded Andromeda XVIII and treated HIZSS 3A \& B as one bound object, assuming a mass ratio of 13:1 \citep{Begum_2005}.




To handle distance uncertainties, we recalculated GRV predictions with each target moved along the line of sight to the 1$\sigma$ upper limit of its observed distance $d_{_{MW}}$ (using the 1$\sigma$ lower limit instead yielded almost identical results).
\begin{eqnarray}
	\sigma_{pos} \equiv \left| GRV_{model} \left(d_{_{MW}} + \sigma_{d_{MW}} \right) - GRV_{model} \left(d_{_{MW}} \right) \right|
	\label{sigma_pos}
\end{eqnarray}

Here, $\sigma_{d_{MW}}$ is the uncertainty in the distance to a target galaxy whose most likely distance is $d_{_{MW}}$. We can now combine HRV measurement uncertainties $\sigma_{v_h}$ with those on GRV predictions caused by distance uncertainties. Thus, the contribution to the $\chi^2$ statistic from any galaxy $i$ is
\begin{eqnarray}
	{\chi_{_i}}^2 &\equiv & \left( \frac{GRV_{model} - GRV_{obs}}{\sigma} \right)^2 ~~\text{ where} \\
	\sigma &=& \sqrt{{\sigma_{pos}}^2 + {\sigma_{v_h}}^2}
	\label{sigma}
\end{eqnarray}

Uncertainty in the distance to M31 has other subtle effects on our analysis. The gravitational field in the LG would be altered if M31 was at a different distance than the assumed 783 kpc. However, we neglect such effects because, towards the edge of the LG, the only relevant factors are the masses of the MW and M31. In any case, the rather small uncertainty in its distance of 25 kpc \citep{McConnachie_2012} is unlikely to affect our results much because the closest target galaxy to M31 is still $\ga 200$ kpc from it (top panel of Figure \ref{Velocity_field_2D_LMC}). Nonetheless, this effect is included directly in our 3D model (Section \ref{3D_method}).

We used a grid method to explore the parameter space spanned by the total MW and M31 mass, the fraction $q_{_{MW}}$ of this in the MW, the LSR speed $v_{c,\odot}$ and the LMC mass. We use uniform priors on model parameters except $v_{c, \odot}$, for which we assume a Gaussian prior of ${239 \pm 5}$ km/s \citep{McMillan_2011} and add a corresponding contribution to the total $\chi^2$. Thus,
\begin{eqnarray}
	\chi^2 ~=~ \left( \frac{v_{c, \odot} - v_{c, \odot \text{,nominal}}}{\sigma_{v_{c, \odot}}} \right)^2 ~+ \sum_{\begin{array}{c} \text{Target}\\ \text{galaxies}\end{array}}{\chi_{_i}}^2
	\label{Chi_sq_total}
\end{eqnarray}

Because $\sigma_{pos}$ varies slightly with the model parameters, our error budgets become model-dependent. Thus, the best-fitting model is not just that which minimises $\chi^2$. We quantify the relative probabilities of different models using
\begin{eqnarray}
	P(\text{Model } | \text{ Observations}) ~\propto~ {\rm e}^{^{-\frac{\chi^2}{2}}} \prod_{i} \frac{1}{\sigma_{_i}}
	\label{P_final}
\end{eqnarray}

We will focus on how observed GRVs deviate from those predicted by our best-fitting model. To facilitate the discussion, we define
\begin{eqnarray}
	\Delta GRV ~\equiv~ GRV_{obs} - GRV_{model}
\end{eqnarray}


\subsection{2D Results} 

\begin{figure}
	\centering 
		\includegraphics [width = 8.5cm] {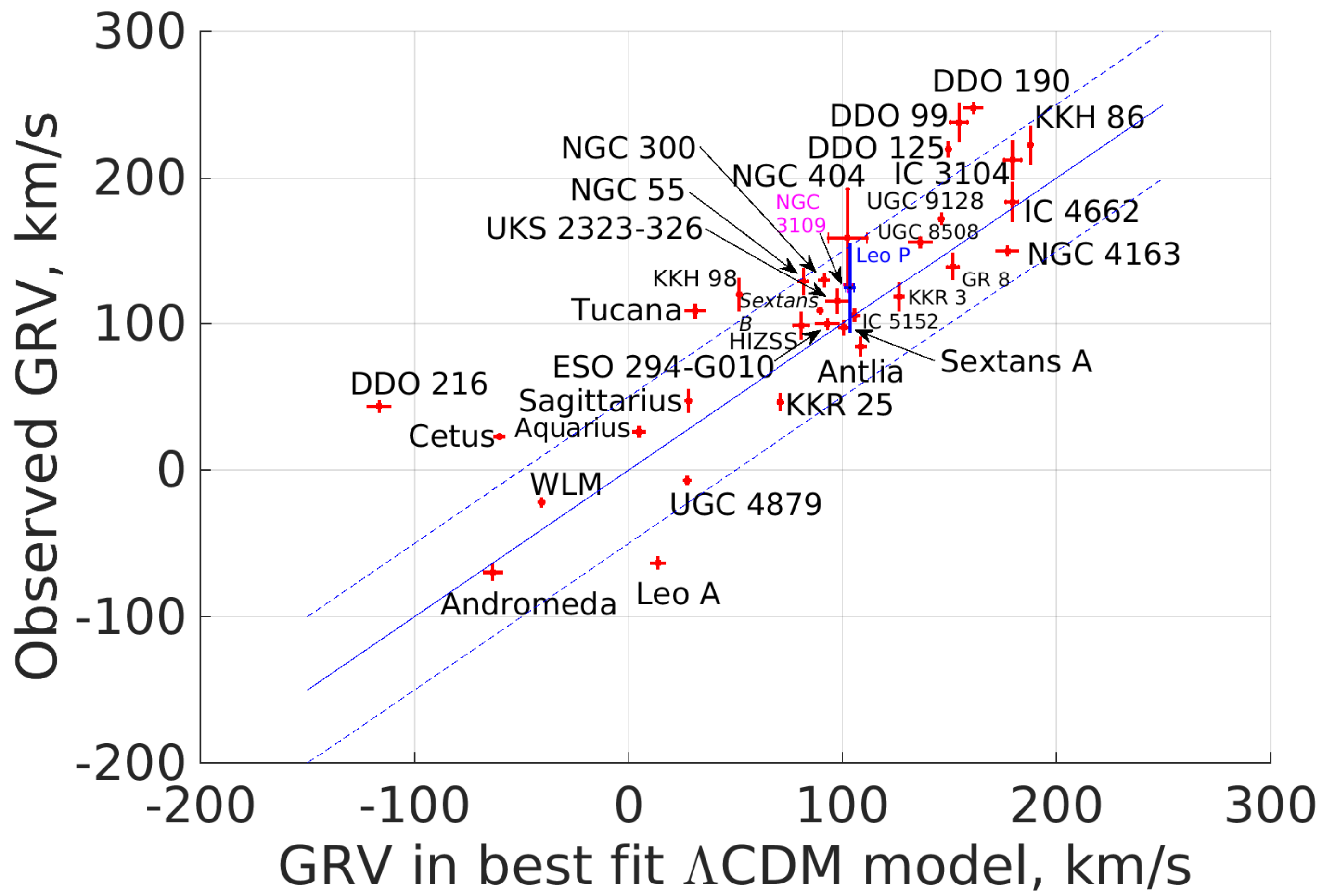}
	\caption{Comparison between predicted and observed GRVs of indicated galaxies in our best-fitting axisymmetric model including Centaurus A and the LMC. The adopted model parameters are given in Table \ref{Best_fit_parameters}. The line of equality is shown in solid blue. Two parallel lines (dashed blue) offset by 50 km/s are also shown. Assuming our model is accurate to $\ssim 25$ km/s, it is unlikely in the context of $\Lambda$CDM to find many galaxies far outside this range. Generally, a larger GRV indicates a larger distance (for reference, Aquarius is $\ssim 1$ Mpc from the LG barycentre).}
	\label{GRV_LCDM_comparison_Cetus}
\end{figure}

In Figure \ref{GRV_LCDM_comparison_Cetus}, we compare model-predicted and observed GRVs in the best-fitting model (parameters given in third column of Table \ref{Best_fit_parameters}).\footnote{This is similar to Figure 8 of \citet{Banik_Zhao_2016}, but the LMC is now included at the best-fitting $q_{_{LMC}}$.} We also show results for Cetus and DDO 216. Despite being quite close to M31, they seem to be in a region with a smooth velocity field (Figure \ref{Velocity_field_2D_LMC}). This allows for a well-defined model prediction. However, the absence of M33 from our model may make these predictions less reliable than for other target galaxies. This is especially true with DDO 216, which is closer to M31. We will return to this point later.

\begin{figure}
	\centering 
		\includegraphics [width = 8.5cm] {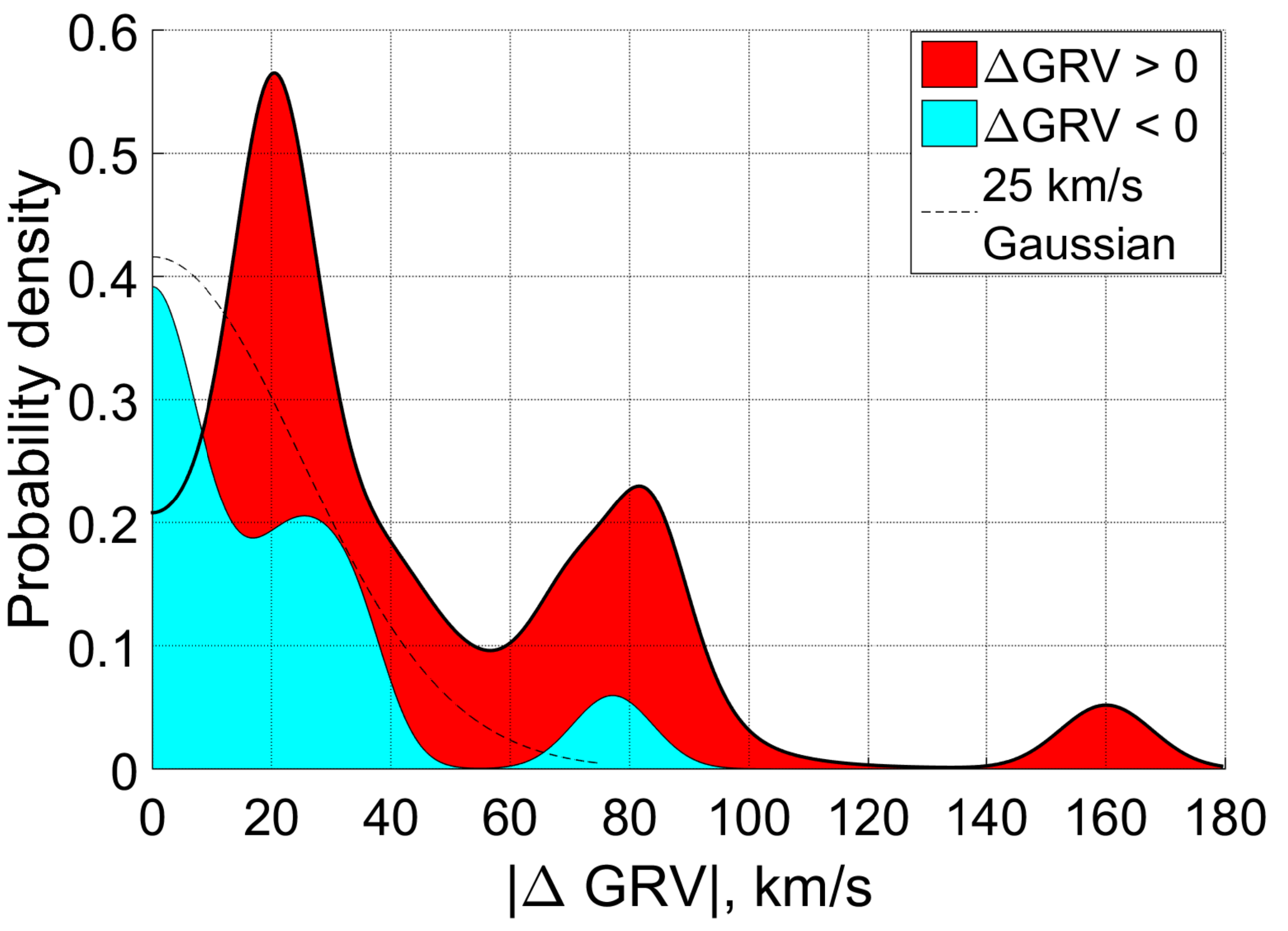}
	\caption{Histogram showing observed $-$ predicted GRVs of our target galaxies using our best fitting 2D model. The area of each square corresponds to 2 galaxies. Each data point was convolved with a Gaussian of width $\sigma = \sqrt{{\sigma_{pos}}^2 + {\sigma_{v_h}}^2 + {\sigma_{v_{c, \odot}}}^2}$. This matches the $\Delta GRV < 0$ subsample (solid blue) quite well, especially when Leo A is excluded as this removes the blue bump near 75 km/s. However, it does not match the $\Delta GRV > 0$ subsample (solid red).}
	\label{Delta_GRV_histogram_LMC_2D}
\end{figure}

There is a tendency for observed GRVs to exceed model predictions (Figure \ref{GRV_LCDM_comparison_Cetus}). To gain a better feel for this phenomenon, we constructed a histogram of all the $\Delta GRV$s. Errors due to distance and HRV uncertainties are accounted for by convolving each data point with a Gaussian of the appropriate width (Equation \ref{sigma}). To account for an uncertain LSR speed (which is only weakly constrained by our investigation), we also added the 5 km/s uncertainty on this in quadrature. The results are shown in Figure \ref{Delta_GRV_histogram_LMC_2D}.

Our model is not a perfect representation of $\Lambda$CDM and can only really be expected to get GRV predictions accurate to $\ssim 25$ km/s \citep{Aragon_Calvo_2011}. Thus, one expects the distribution of $\Delta GRV$s to be broadly consistent with a Gaussian of this width. Indeed, this appears to be the case for galaxies with $\Delta GRV < 0$ (blue in Figure \ref{Delta_GRV_histogram_LMC_2D}).

On the other hand, this is not true for galaxies which have $\Delta GRV > 0$. One can dismiss the bump in the histogram near 160 km/s due to DDO 216 on the grounds that it may be too close to M31 and thus the velocity field may be disturbed there. This is not the case in our model (Figure \ref{Velocity_field_2D_LMC}) but one can envisage that it is true in the real world when one considers additional effects e.g. interactions with massive M31 satellites such as M33. However, it is very difficult to dismiss the bump at 80 km/s in this way because it corresponds to several galaxies, some of which are quite far from the LG (top panel of Figure \ref{Distance_GRV_correlation}). The presence of this feature along with the expected bump near ${\Delta GRV = 0}$ suggests the existence of some additional process responsible for a few galaxies having GRVs much higher than expected in our model. We consider some possible solutions to this high velocity galaxy problem in Section \ref{Discussion}.

\begin{figure}
	\centering 
		\includegraphics [width = 8.5cm] {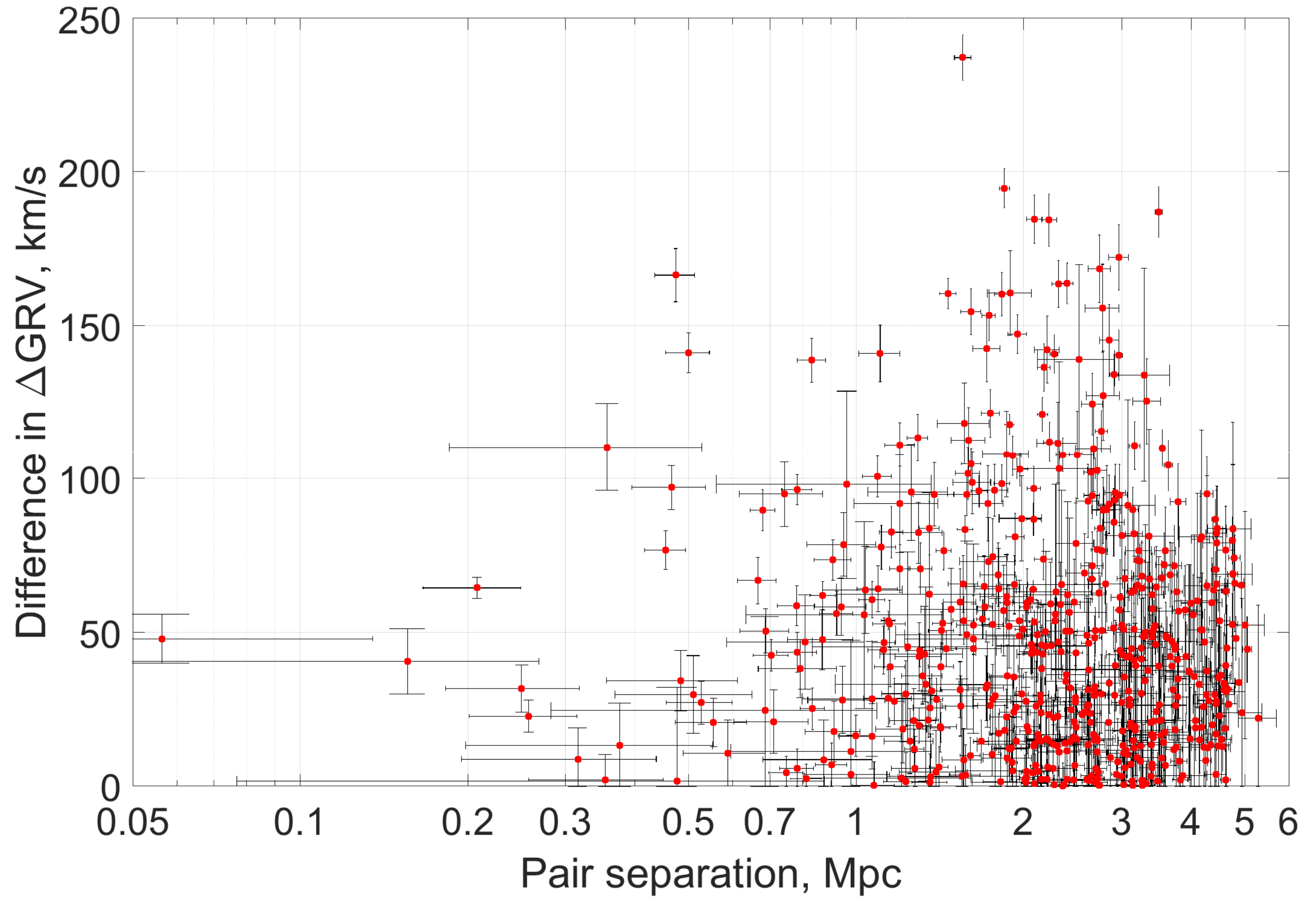}
	\caption{Distances between galaxies in our sample are shown here against the difference between their $\Delta GRV$s. The errors on these correspond to errors on measured distances and radial velocities of both galaxies being compared. For galaxies near each other, model predictions for their GRVs are similar such that the $y$-axis essentially just shows the difference between their observed HRVs. Sometimes, these differences are large even for nearby galaxies. These cases appear towards the upper left potion of this figure. The most striking examples are listed in Table \ref{Two_stream_investigation_results}.}
	\label{Pair_comparison_2D}
\end{figure}

At the positions of our target galaxies, we expect the velocity field of the LG to be smooth (Figure \ref{Velocity_field_2D_LMC}). To see if this is the case, we determined the distance between each pair of targets and the difference in their $\Delta GRV$s (Figure \ref{Pair_comparison_2D}).\footnote{Errors on mutual separations are over-estimated because we add distance errors in quadrature.}


Some examples are apparent where galaxies are quite near each other but have a very different $\Delta GRV$. In these situations, because model predictions should be very similar, the difference in $\Delta GRV$s must be mostly due to a difference in observed HRVs.\footnote{We found it helpful to use $\Delta GRV$s instead of directly observed HRVs as this allows us to include effects such as galaxies further away along the same line of sight being expected to have a higher HRV.} More information is given about some of these cases in Table \ref{Two_stream_investigation_results}.

A few such discrepant pairs are expected given that there are 561 pairs in total. However, the magnitude of the difference between $\Delta GRV$s is rather large in some cases, suggesting that the velocity field in the LG may not be as smooth as in our model. The most convincing examples of this are related to the galaxy NGC 4163, without which the case for a disturbed velocity field is greatly weakened. The galaxies near it (DDO 99, 125 and 190) all seem to have much higher $\Delta GRV$s, suggesting that perhaps the GRV of NGC 4163 is unusually low. In fact, it has the third-lowest $\Delta GRV$ of ${-26.9 \pm 7.6}$ km/s. This may not seem very low, but we will see later that a more detailed 3D model of the LG predicts a much higher GRV for this galaxy than is observed (Section \ref{3D_method}).

\begin{table}
	\begin{tabular}{  m{1.35cm}  m{1.35cm} m{1.25cm}  m{1.25cm}  m{1.2cm} }  
	\hline
	Galaxy 1&Galaxy 2&Separation&Ratio&Difference\\
	& & (Mpc) & &in $\Delta GRV$\\
	& & & & (km/s) \\
	\hline
	DDO 99&NGC 4163&0.36$\pm$0.17&0.13$\pm$0.06&110.4$\pm$14\\
	DDO 125&NGC 4163&0.47$\pm$0.07&0.17$\pm$0.03&97.1$\pm$7.3\\
	\hline
	KKR 3&DDO 190&0.74$\pm$0.13&0.30$\pm$0.05&95.0$\pm$10.5\\
	Andromeda&DDO 216&0.47$\pm$0.04&0.56$\pm$0.05&166.5$\pm$8.6\\
	WLM&DDO 216&0.50$\pm$0.05&0.54$\pm$0.05&141.3$\pm$6.5\\
	UGC 8508&DDO 190&0.67$\pm$0.05&0.25$\pm$0.02&66.9$\pm$7.6\\
	WLM&Cetus&0.21$\pm$0.04&0.25$\pm$0.05&64.5$\pm$3.4\\
	\hline
	\end{tabular}
	\caption{Differences between the $\Delta GRV$ of nearby galaxies are shown here, with our 2D model used to obtain predicted GRVs. The error budgets account for uncertainties in HRV and distance measurements of both targets. Their separation is shown in physical units and as a fraction of the mean of the distances from the MW to each of them. We show only the most extreme examples of galaxies near each other but with a very different $\Delta GRV$ (most convincing examples near top). Results for all galaxy pairs are shown in Figure \ref{Pair_comparison_2D}.}
\label{Two_stream_investigation_results}
\end{table}

A 3D model allows us to test our conclusions more rigorously by directly including many more objects, several of which are quite far from the MW$-$M31 line. The inclusion of the LMC and M33 can help to make our model more reliable closer to the MW and M31, respectively. This is one reason why we felt comfortable adding Cetus and DDO 216 to our sample. Our model should also be more reliable further from the LG as it now includes more of the most massive objects just outside it. However, we do not take advantage of this by expanding our sample outwards.

\begin{table}
	\begin{tabular}{  m{1.15cm}  m{2.95cm} m{1.1cm}  m{1.32cm}}
	\hline
	\multicolumn{1}{l}{Parameter} & Meaning \& units & Best- & Best- \\
	 &   & fitting & fitting \\
	& & value & value \\	 
	& & in 2D &  in 3D \\ \hline
	$M$ & LG mass, ${10}^{12} M_\odot$ & 2.756 & 4.088 \\
	$q_{_{MW}}$ & $\frac{M_{_{MW}}}{M}$ & 0.356 & 0.497 \\
	$q_{_{LMC}}$ & $\frac{M_{_{LMC}}}{M_{_{MW}}}$ & 0.157 & 0.099 \\
	$v_{c, \odot}$ & LSR speed, km/s & 239 & 223.0 \\ \cline{1-3}
	$v_{f, M31}$ & $v_{_f}$ of M31, km/s & \multicolumn{1}{l|}{225} & 240.3 \\	
	$d_{_{M31}}$ & Distance to M31, kpc & \multicolumn{1}{l|}{783} & 707 \\
	$M_{\text{Cen A}}$ & Cen A mass, ${10}^{12} M_\odot$ & \multicolumn{1}{l|}{4} & 5.883 \\ [5pt] \cline{4-4}
	$U_\odot$ & Components of the & 14.1 & 11.1\\
	$V_\odot$ & non-circular motion of & 14.6 & 12.2\\
	$W_\odot$ & Sun in the MW, km/s & 6.9 & 7.2\\
	\hline
	$H_{_0}$ & Hubble constant & 67.3 & 70 \\ 
	$\Omega_{m,0}$ & Present matter density & 0.315 & 0.27\\
	& in the Universe $\div \frac{3{H_{_0}}^2}{8 \rm{\pi} G}$ & & \\ [5pt]
	\hline
	\end{tabular}
	\caption{The parameters of our best-fitting axisymmetric (2D) and 3D models are given here. $q_{_{LMC}}$ is defined in Equation \ref{LMC_adjustment}. The top section of this table contains the parameters we varied using a grid search in our 2D model \citep{Banik_Zhao_2016} or using a gradient descent method in 3D (Section \ref{3D_statistical_analysis}). The central section contains the parameters associated with the non-circular motion of the Sun in the Milky Way, which we obtain from \citet{Francis_2014} for the 2D model and from \citet{Schonrich_2010} for the 3D model. This section also contains two parameters related to M31. In the 2D model, its distance estimate is from \citet{McConnachie_2012} while the 3D model uses a prior of ${770 \pm 40}$ kpc \citep{Ma_2010}. Its rotation curve flatlines at a level $v_{f, M31}$ which is fixed in the 2D model but has a prior of ${240 \pm 10}$ km/s in the 3D model \citep{Carignan_2006}. This model assumes $v_{_f} = v_{c,\odot}$ for the MW whereas the 2D model fixes the former at 180 km/s \citep{Kafle_2012} and uses a prior on the latter of ${239 \pm 5}$ km/s \citep{McMillan_2011}. We adopt a flat dark energy-dominated cosmology whose parameters are fixed at values given in the bottom section, with the 2D results based on those of \citet{Planck_2015} while the 3D results are based on \citet{Komatsu_2011}. Both models start when the cosmic scale-factor ${a = 0.1}$.}
\label{Best_fit_parameters}
\end{table}

\section{The 3D Method}
\label{3D_method}

\subsection{Governing equations}

The 3D algorithm we employ is explained in more detail in Appendix A of \citet{Shaya_2011}, which applies the numerical action method to solve the governing equations. A more detailed attempt was later made to use this method to understand the dynamics of LG galaxies \citep{Peebles_2013}. We briefly review some of the key aspects of how the model works.

We adapted a previous dynamical model of the LG based on the objects included in \citet[][Table 1]{Shaya_2011}. This brightness-based catalogue is similar to the massive galaxies used in our analysis (Table \ref{Massive_galaxy_list_3D}). However, it misses the vast majority of the galaxies analysed in \citet{Banik_Zhao_2016}, which is a major shortcoming because LG dwarfs $\ssim 1-3$ Mpc away turned out to be crucial to its conclusions. Thus, if not already present in our 3D model, we added the galaxies analysed in that work to it as test particles satisfying the equation of motion
\begin{eqnarray}
	{\overset{..}{\bm r} } ~&=&~ {H_{_0}}^2\Omega_{_{\Lambda, 0}}{\bm r} ~- \sum_{\begin{array}{r} \text{j = Distant}\\ \text{massive}\\ \text{particles}\end{array}} \frac{G M_j \left( \bm r - \bm r_{_j} \right)} {|\bm r - \bm r_{_j} |^3} \nonumber \\
	&~&- \sum_{\begin{array}{r} \text{j = Nearby}\\ \text{massive}\\ \text{particles}\end{array}} \frac{G M_j \left( \bm r - \bm r_{_j} \right)  \left( {r_{_c}}^2 + {r_{_{S,j}}}^2 \right)} {\left( |\bm r - \bm r_{_j} |^2 + {r_{_c}}^2 \right) {r_{_{S,j}}}^3 }\nonumber \\
	\label{Equation_of_motion_3D}
\end{eqnarray}

When determining the force between any pair of massive galaxies, the value of $r_{_{S}}$ used corresponds to the galaxy with the larger $r_{_{S}}$. The massive galaxies in this analysis are given in Table \ref{Massive_galaxy_list_3D}. The distances and HRVs shown are best-fitting values obtained based on trying to match all observational constraints within their uncertainties (Section \ref{3D_statistical_analysis}).

The gravitational field near massive particles is handled slightly differently than in our 2D model. For any given test particle $A$, an explicit distinction is now drawn between massive particles whose $r_{_{S}}$ is below the distance to $A$ and masses for which this is not the case, forces from which are handled using a pure inverse square law. Forces from nearby masses at first rise linearly with separation before falling as $F \propto \frac{1}{r}$, recovering the observed flat rotation curves of galaxies. The transition occurs around $r_{_c} = 10$ kpc.

For the MW and M31, we define $r_{_{S}}$ in the same way as previously, though we now add the assumption that the LSR speed is the same as $v_{_f}$ for the MW. Its value is allowed to float, with a prior assumption of ${240 \pm 10}$ km/s. We use the same value for M31. For other massive galaxies, we assume $r_{_{S}} = 100$ kpc to avoid an adjustment each time their masses are altered.

Some differences with Equation \ref{Equation_of_motion_2D} are apparent. The part of the cosmological acceleration $\ddot{a}$ caused by dark energy is handled in the same way but the part caused by matter is not. Previously, we treated the Universe as homogeneous except for a few massive particles. This meant that, without these particles, we would need to recover the cosmic expansion $\bm r \propto a$, which is only possible if $\ddot{\bm{r}} = \frac{\ddot{a}}{a}\bm r$.

Here, we treat the Universe as empty except for the massive particles that we explicitly include. Because the Universe is homogeneous on large scales, an accurate understanding of all the mass interior to a sufficiently distant test particle also leads to its separation from us changing with time as $\bm r \propto a$. To see if this applies to our model, we determined how much mass was in our simulation out to the distance of M101, the most distant galaxy in our sample. The result of $4.9 \times 10^{13} M_\odot$ corresponds to a sphere of radius 7.01 Mpc filled with matter at a density equal to the present cosmic mean value. This is similar to the observed distance of M101 \citep{Shappee_2011}, suggesting that the massive galaxies in our model mimic a smooth distribution on large scales with the correct density. A similar conclusion would be reached if we only consider galaxies within 3 Mpc of the MW.



\begin{table}
	\begin{tabular}{lcrc}
	\hline
	Galaxy & Distance, & HRV, & Mass, \\
	& Mpc & km/s & $10^{12} M_\odot$ \\
	\hline
	Milky Way & 0.008 & $-$11.10 & 1.8302 \\
	Andromeda (Messier 31) & 0.707 & $-$309.18 & 2.0567 \\
	Centaurus A & 3.736 & 504.52 & 5.8831 \\ \hline
	Messier 101 & 7.391 & 439.62 & 9.3108 \\
	Messier 94 & 4.366 & 324.31 & 8.8144 \\
	Sculptor & 4.095 & 246.97 & 6.9296 \\
	NGC 6946 & 5.859 & 107.38 & 4.6142 \\
	Messier 81 & 3.625 & 73.48 & 4.0625 \\
	Maffei & 3.988 & $-$28.75 & 3.4924 \\
	IC 342 & 3.350 & $-$12.98 & 1.2994 \\
	Triangulum (Messier 33) & 0.948 & $-$192.72 & 0.2214 \\
	Large Magellanic Cloud & 0.065 & 235.97 & 0.2007 \\
	NGC 55 & 2.035 & 163.16 & 0.1323 \\
	NGC 300 & 1.963 & 158.70 & 0.1073 \\
	IC 10 & 0.781 & $-$338.02 & 0.0437 \\
	NGC 185 & 0.706 & $-$213.37 & 0.0129 \\
	IC 5152 & 1.878 & 138.56 & 0.0094 \\
	NGC 147 & 0.679 & $-$201.04 & 0.0064 \\
	NGC 6822 & 0.510 & $-$69.93 & 0.0059 \\
	\hline
	\end{tabular}
	\caption{Data on the massive galaxies in our 3D model using a similar catalogue to \citet[][Table 1]{Shaya_2011}. Distances and masses are allowed to vary to best match observations, though their prior distributions are not uniform (see text). The masses derived in our model correspond to the total halo mass of each system, some of which is located beyond its virial radius \citep{Fattahi_2016}. The top section of this table contains galaxies which are also directly included as massive extended objects in our 2D model (Section \ref{2D_review}). The remaining galaxies are sorted in descending order of simulated mass. For clarity, we abbreviated the names of galaxies from the New General Catalogue (NGC) and Index Catalogue (IC).}
\label{Massive_galaxy_list_3D}
\end{table}

The equations of motion are solved by adjusting a trial trajectory towards the true one. An incorrect trajectory will have a mismatch between the acceleration along it and that expected due to the gravity of other particles. Thus, at each timestep, the positions of all the particles are adjusted to try and equalise the gravitational field acting on each one with the acceleration $\overset{..}{\bm r}$ it experiences along its trajectory. This is done assuming both respond linearly to a position adjustment, although only the latter does. Thus, a solution can only be obtained after several iterations, each of which is reliant on a matrix inversion to handle the highly inter-connected nature of the problem. Certain shortcuts are taken for test particles because their position has no effect on the forces felt by other particles.

This method of solution is second-order accurate because of the standard finite differencing scheme used to obtain accelerations from a series of discrete positions valid at known times. Due to the large number of particle pairs, an adaptive timestep scheme is impractical. Instead, we adapt the temporal resolution to the problem in a fixed way based on physical considerations. Each timestep corresponds to an equal increment in the cosmic scale-factor $a$. We use 500 steps between when $a = 0.1$ and the present time ($a \equiv 1$).

To check if we have adequate resolution, the problem is solved using forward integration instead with 5000 timesteps equally spaced in $a$. The maximum error in the present position is 0.23 kpc while that in the velocity is 0.84 km/s. Both errors are very small, suggesting that we have enough resolution. Some other checks are also done to verify the numerical accuracy of our solution \citep[][Section 2.4]{Shaya_2011}.

\begin{figure}
	\centering 
		\includegraphics [width = 8.5cm] {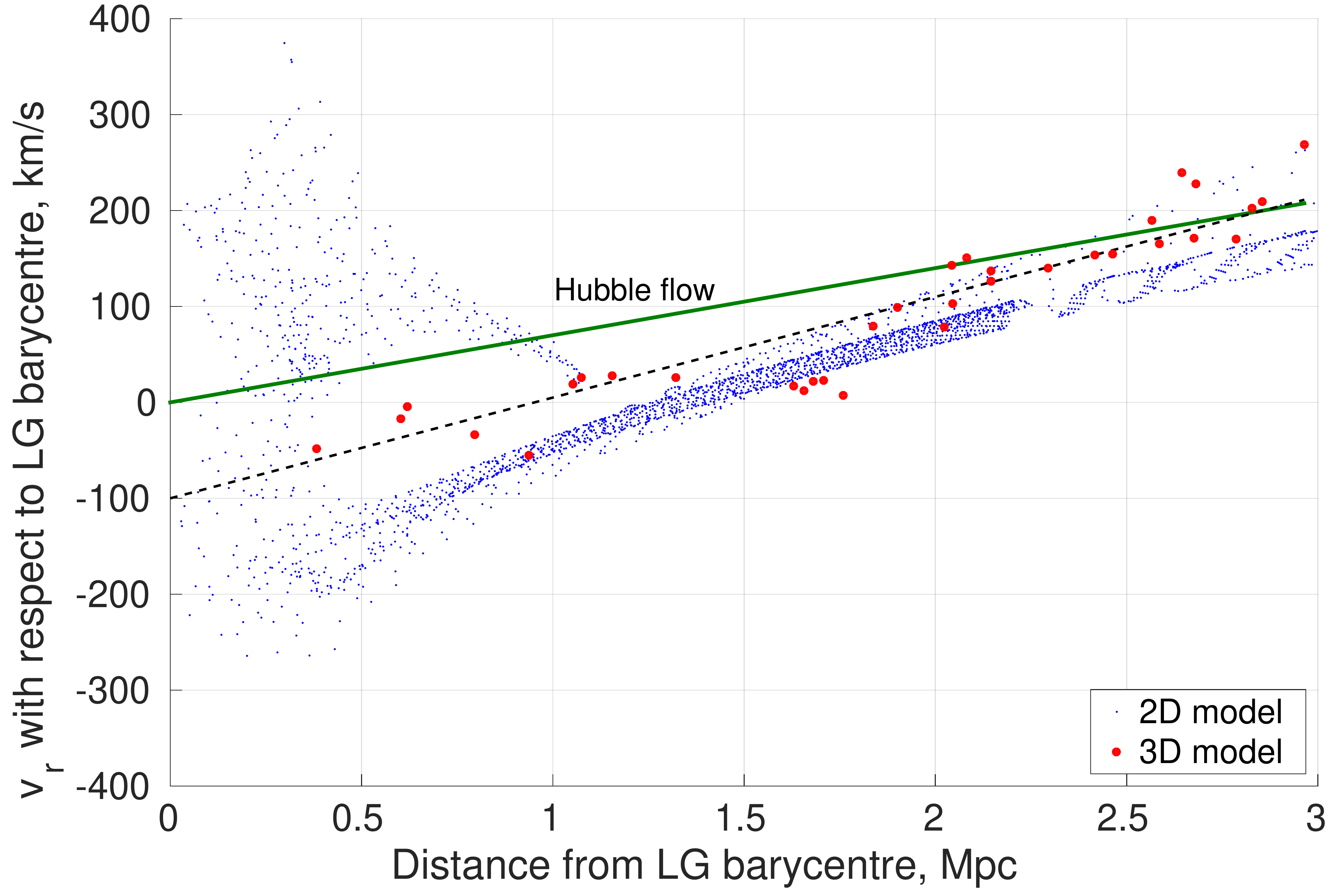}
	\caption{Radial velocities of test particles with respect to the LG barycentre are shown in blue for our 2D model with parameters matched to our best-fitting 3D model (Table \ref{Best_fit_parameters}), results of which are shown as large red dots. The solid green line is the Hubble flow relation for $H_{_0} = 70$ km/s/Mpc, our adopted value. The dashed black line has a gradient ${1.5\times}$ larger. Due to the effect of gravity, it provides a better fit to the 3D model within the LG.}
	\label{LG_Hubble_diagram_3D}
\end{figure}


The Hubble diagram for our best-fitting 3D model is shown in Figure \ref{LG_Hubble_diagram_3D}. For comparison, we overlay results from an axisymmetric model using the same model parameters as this 3D model (last column of Table \ref{Best_fit_parameters}). The basic trend of increasing radial velocity with distance is apparent in both models, even though they are constructed quite differently.



\subsection{Statistical analysis}
\label{3D_statistical_analysis}

Like our axisymmetric model, our 3D model accurately matches the observed sky positions of target galaxies. However, this is achieved rather differently. Instead of integrating the equations of motion forwards in time and using the Newton-Raphson method to very precisely match the present position, the 3D model integrates backwards in time starting from a position along the line of sight towards a target galaxy. We no longer require agreement between simulated and observed heliocentric distances. Instead, we add a contribution to the total $\chi^2$ of the model if there is a mismatch. Handling distance uncertainties in this way makes error budgets model-independent, reducing the determination of relative model likelihoods to a simple comparison of their $\chi^2$ statistics.

The distance errors $\sigma_{d}$ come from observations. For M31, we use a slightly closer and more uncertain estimate \citep[${770 \pm 40}$ kpc,][]{Ma_2010}. Galaxies outside the LG might be affected by objects beyond the region covered by our analysis. It can also be difficult to determine the mass ratios between galaxies in an extended group and thus the location of its centre of mass. For these reasons, we use a fairly large uncertainty for such distant objects.
\begin{eqnarray}
	\frac{\sigma_{d}}{d_{_{MW}}} ~=~ \frac{1}{10}~~\text{ if } d_{_{MW}} > 3.2~\text{Mpc}
	\label{Distance_uncertainty_perturbers}
\end{eqnarray}

Mismatches between observed and simulated GRVs are handled similarly, based on a tolerance of 20 km/s rather than the actual HRV measurement uncertainty. This is because we do not expect our model to be much more accurate as a representation of $\Lambda$CDM considering the level of scatter about the Hubble flow in more detailed simulations \citep{Aragon_Calvo_2011}. As $\sigma_{v_h}$ is always much smaller than this, the effect of raising it to 20 km/s is similar to adding an extra 20 km/s dispersion term to Equation \ref{sigma}. This prevents the model placing undue statistical weight on a galaxy with very precise observations, given that the model itself also has uncertainties.


We made use of proper motion data for M31, M33, the LMC, IC10 and Leo I. This was done by adding a penalty to $\chi^2$ when simulated and observed values disagree, with observational error estimates taken at face value.

Unlike in our 2D model, Equation \ref{Initial_conditions} is no longer strictly enforced at the start of our simulations because this is difficult to achieve when integrating backwards. Instead, we penalise models which fail to enforce it.
\begin{eqnarray}
	\Delta \chi^2 ~=~ \frac{| \overbrace{\bm{v_{_i}} - H_{_i} \bm{r}_{_i}}^{\bm{v}_{pec} \left( t = t_i \right)} |^2}{{\sigma_{_v}}^2}
	\label{chi_sq_contribution_v_pec}
\end{eqnarray}

\begin{figure}
	\centering 
		\includegraphics [width = 8.5cm] {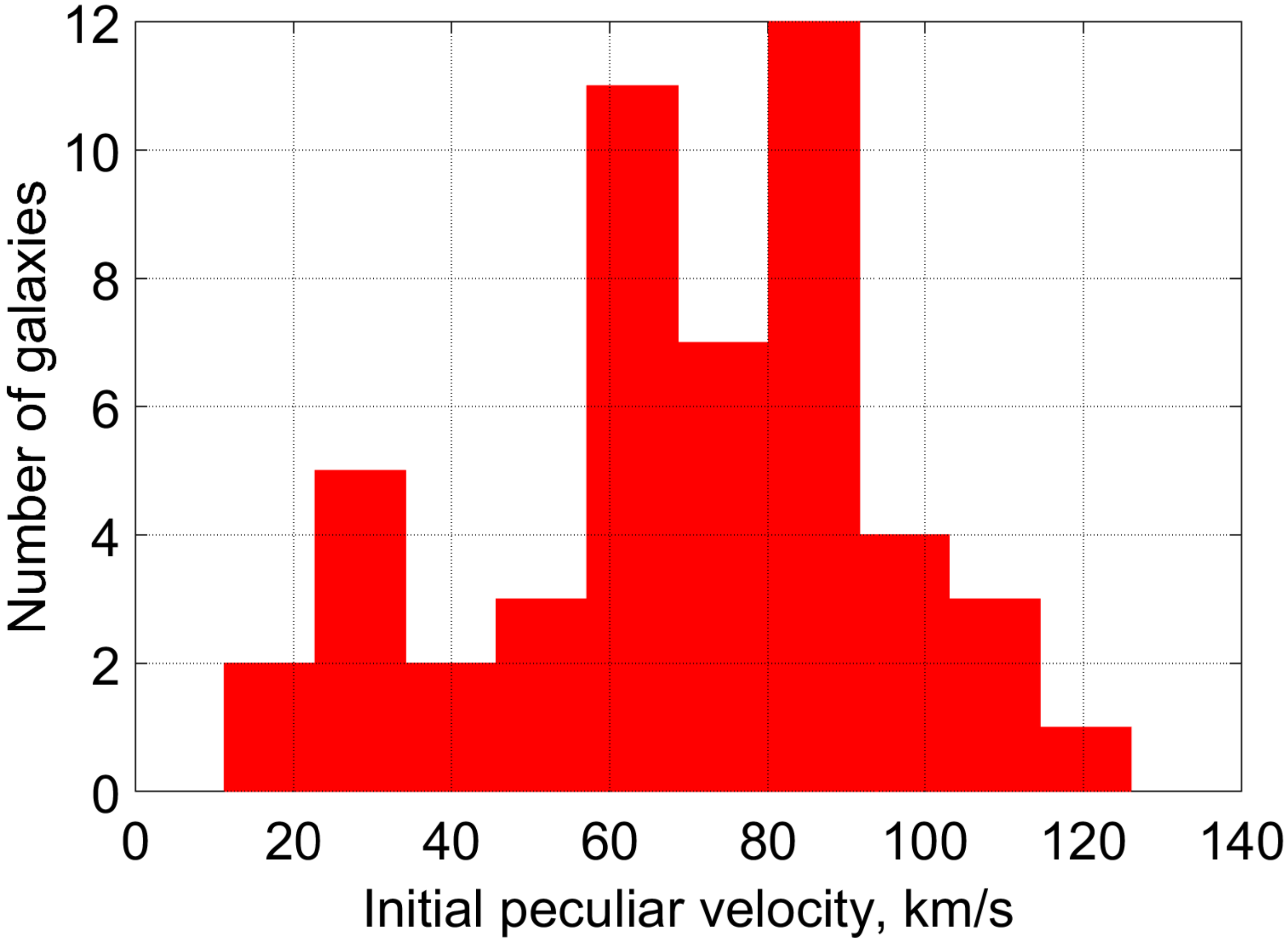}
	\caption{Histogram showing deviations from the Hubble flow (Equation \ref{Initial_conditions}) at the start of our best-fitting 3D simulation. We allow a tolerance of 50 km/s (Equation \ref{chi_sq_contribution_v_pec}). The mean peculiar velocity then was 69 km/s, with a $1\sigma$ confidence interval of 38$-$91 km/s.}
	\label{v_pec_histogram}
\end{figure}

We assume that the typical peculiar velocity $\bm{v}_{pec}$ when $a = 0.1$ was $\sigma_{_v} = 50$ km/s based on present-day deviations from the Hubble flow (Figure \ref{LG_Hubble_diagram_3D}). This is a 1D measure which underestimates typical values of ${v}_{pec}$ today. However, the nearly homogeneous state of the Universe at recombination \citep{Planck_2015} implies that ${v}_{pec}$ is typically larger at the present time than when our simulations started. The distribution of ${v}_{pec}$ at that time is shown in Figure \ref{v_pec_histogram} for our best-fitting 3D model. All 50 galaxies in this model are represented, some of which are outside the LG.

We do not fix the masses of any of our simulated galaxies which are treated as massive (Table \ref{Massive_galaxy_list_3D}). However, the prior we use prefers a particular value based on assuming a mass-to-light ratio of 50 times the Solar value in the near-infrared K-band \citep{Tully_2013}. Observational estimates of the luminosity in this band $L_K$ are based on a particular distance to each target. If e.g. the distance in the model is less, then the model implies that the target is likely closer to us and thus intrinsically fainter for the same apparent magnitude. This makes it likely to be less massive. Accounting for this, we define the preferred mass estimate
\begin{eqnarray}
	M_c ~\equiv~ 50 L_K \left( \frac{M}{L_K} \right)_\odot \left( \frac{d_{model}}{d_{obs}} \right)^2
	\label{M_c}
\end{eqnarray}

Using a different mass $M$ incurs an extra $\chi^2$ cost of
\begin{eqnarray}
	\Delta \chi^2 ~=~ \left[ \frac{Ln \left( \frac{M}{M_c}\right)}{Ln ~1.5} \right]^2
	\label{chi_sq_contribution_M}
\end{eqnarray}

For the MW and M31, a slightly different procedure is used. There is no a priori preference towards any particular mass for either galaxy, but a particular \emph{ratio} between their masses is preferred. This is the ratio of their values of $M_c$.
\begin{eqnarray}
	\Delta \chi^2 ~= \left( \frac{Ln~\frac{M_{MW}}{M_{M31}} ~-~ Ln~\frac{M_{MW,c}}{M_{M31,c}}}{Ln~1.25}\right)^2
	\label{chi_sq_contribution_MW_M31_ratio}
\end{eqnarray}

Our model now has too many parameters to permit a grid search through them. Thus, we only present results from our best-fitting 3D model. This is obtained by minimising $\chi^2$ using a downhill-seeking walk through parameter space \citep[][Section 2.2]{Shaya_2011}. Each parameter $A$ is varied by a small amount $\delta A$ in an attempt to reduce $\chi^2$. If this does not happen, then the algorithm restores the previous solution and sets
\begin{eqnarray}
	\delta A ~\to ~-\frac{1}{2} \delta A ~~ \left( \chi^2 \text{ increased} \right)
\end{eqnarray}

As well as reversing the sign of $\delta A$, it is important to reduce its magnitude because the increase in $\chi^2$ is often caused by overshooting its minimum with respect to $A$.

When we are fortunate in that a parameter adjustment reduces $\chi^2$, we accelerate the convergence by setting
\begin{eqnarray}  
	\delta A ~\to~ \frac{5}{4} \delta A ~~~~~ \left( \chi^2 \text{ decreased} \right)
\end{eqnarray}


To avoid the parameter adjustments being too large or too small, a cap and floor are imposed on $\left| \delta A \right|$ such that
\begin{eqnarray}
	10^{-5} ~<~ \left| \frac{\delta A}{A} \right| ~<~ 10^{-1}
\end{eqnarray}


\section{3D Results}
\label{3D_results}

We started with a solution to the equations of motion which best matched available constraints on the massive galaxies in our analysis (Table \ref{Massive_galaxy_list_3D}). We then added test particles one at a time, giving the algorithm an opportunity to adjust the masses and trajectories of the massive objects in order to better accommodate observational constraints from the corresponding LG dwarf. Once we had included our complete sample, we varied all the parameters one at a time to see if we could achieve any further reduction in $\chi^2$. After repeating this a few times, it became clear that the preferred solution was not changing. Some of the most important parameters associated with this model are shown in Table \ref{Best_fit_parameters}.

The optimal value of $q_{_{MW}}$ is 0.50, slightly higher than the ${0.36 \pm 0.04}$ preferred by our axisymmetric analysis \citep[error budgets estimated from][Figure 18]{Banik_Zhao_2016}. This is because our 3D model puts M31 at a distance of only 707 kpc (Table \ref{Massive_galaxy_list_3D}), much less than the most likely distance of 783 kpc \citep{McConnachie_2012}. If the model is to be trusted, then the known apparent magnitude of M31 combined with a closer distance implies that it is intrinsically fainter. This reduces the preferred mass of M31 (Equation \ref{M_c}). If we scale up the mass of M31 by $\left(\frac{783}{707} \right)^2$, then $q_{_{MW}}$ would fall to 0.45. Because we include the LMC as part of the MW when determining $q_{_{MW}}$, it seems reasonable to treat M33 as part of M31 rather than as a separate object. Doing so reduces $q_{_{MW}}$ a further $\ssim 0.03$. This makes it consistent with our axisymmetric analysis, assuming both yield similarly uncertain estimates of $q_{_1}$.

We consider several galaxies close enough to M31 for the flatline level of its rotation curve to make some difference. This is especially true with NGC 147 and NGC 185. To a lesser extent, it is also the case for IC 10. This is interesting in light of its measured proper motion \citep{Brunthaler_2007}. Thus, our model may be able to constrain $v_{_{f, M31}}$. It prefers a value of 240.29 km/s, very close to the 240 km/s at which the prior distribution peaks (Table \ref{Best_fit_parameters}). We are unable to determine the precision with which our model constrains this parameter. We believe the suspiciously good agreement (within $0.03\sigma$ of the prior) indicates that our analysis is simply unable to obtain meaningful constraints on $v_{_{f, M31}}$, such that its prior is the most important consideration.

Our analysis preferred a low value of $v_{c, \odot}$, so we focused on adjusting only this parameter to better constrain its optimal value. This did not alter our results. Our 3D analysis alone must prefer even lower values than the best fit, which also considers our prior of ${240 \pm 10}$ km/s. Given that analyses such as these typically constrain $v_{c, \odot}$ to within no better than $\ssim 15$ km/s \citep[e.g.][]{Jorge_2014, Banik_Zhao_2016}, it appears that there is some tension between the 223 km/s preferred by our analysis and the independent\footnote{not based on the timing argument} estimate of ${239 \pm 5}$ km/s \citep{McMillan_2011}. Interestingly, a more recent estimate preferred a lower value of ${232.8 \pm 3.0}$ km/s \citep{McMillan_2016}. However, the reduced uncertainty leads to the same conclusion.

Our model also contains some galaxies quite close to the MW, making it important to have an accurate force law within its virial radius. We may be failing to achieve this by assuming ${v_{c, \odot} = v_{_f}}$, the speed at which the rotation curve of the MW flatlines. It is possible that the best-fitting LSR speed obtained by our algorithm has been dragged down because $v_{_f} \ll v_{c, \odot}$, as suggested by \citet{Kafle_2012}. We hope to relax the assumption that ${v_{c, \odot} = v_{_f}}$ in a future investigation.

Our axisymmetric analysis had almost no preference for a LSR speed different to the 239 km/s peak of its prior distribution \citep[][Table 2]{Banik_Zhao_2016}. The lower value of $v_{c, \odot}$ preferred by our 3D analysis affects $M$, the inferred total mass of the MW and M31. This is because M31 is almost directly ahead of the Sun in its orbit around the MW. Thus, a lower LSR speed implies more of the observed blueshift of M31 must be due to it moving towards the MW, requiring a higher combined mass for these galaxies. We estimate that a 16 km/s reduction in $v_{c, \odot}$ increases $M$ by ${\ssim 0.8\times 10^{12} M_\odot}$ \citep[][Figure 7]{Banik_Zhao_2016}. A higher $M$ is likely also required to counteract the stronger effect of tides raised by Cen A due to its higher inferred mass. Its location very close to the MW-M31 line and relative proximity make it an important consideration. Additionally, we expect a similar effect due to the different assumptions concerning the background density of matter in the LG. Our 2D analysis assumed the LG was filled with matter at the cosmic mean density. In our 3D model, it is treated as empty apart from a few discrete objects like the MW and M31. Our previous results suggest that this should increase the best-fitting value of $M$ by ${\ssim 10^{12} M_\odot}$ \citep[][Section 4.1]{Banik_Zhao_2016}. For these reasons, it is not too surprising that our 3D analysis prefers a higher $M$, even though this is counteracted slightly by the lower preferred distance to M31.


We conducted versions of our axisymmetric analysis without Cen A and with it included at a mass of $4\times10^{12}M_\odot$. The latter provided a much better fit to observations \citep[][Figure 14]{Banik_Zhao_2016}. Our 3D algorithm starts off with Cen A having $M_c = 1.06\times10^{13}M_\odot$ but the analysis prefers a value of just over half this (not problematic given the fairly broad mass priors $-$ see Equation \ref{chi_sq_contribution_M}). This suggests the possibility of constraining the masses of galaxies just outside the LG based on their tidal effect within it. Such an analysis is likely to face degeneracies between masses of different galaxies along a similar line of sight, but it might still be worthwhile.


A comparison between our best-fitting 2D and 3D models is complicated somewhat by the latter having many more degrees of freedom. In particular, it is not required to match the observed distances of LG galaxies. This allows it to place a galaxy further away than observed, increasing its predicted GRV and better explaining a very high observed GRV. We handle this by applying a correction to model-predicted GRVs if they correspond to a simulated galaxy at a different distance than the real one it is supposed to represent. Thus, we set
\begin{eqnarray}
	\label{GRV_adjustment_distance_error}
	GRV_{model} &\to&  GRV_{model} + \left( d_{obs} - d_{model}\right) \alpha H_{_0} \\
	\alpha &\equiv& \frac{1}{H_{_0}}\frac{dv_{_r}}{dr}
\end{eqnarray}

We use $\alpha = 1.5$ because this seems to provide a reasonable description of how radial velocities $v_{_r}$ depend on distances within the LG (Figure \ref{LG_Hubble_diagram_3D}). At long range, we would get $\alpha = 1$. Within the LG, gravity from the MW and M31 is important. Thus, an object further from them has been decelerated less by their gravity. This means that its radial velocity will be higher by a greater amount than in a homogeneously expanding Universe. Neglecting projection effects (which become small a few Mpc from the LG), we see that $\alpha$ should slightly exceed 1.

\begin{figure}
	\centering 
		\includegraphics [width = 8.5cm] {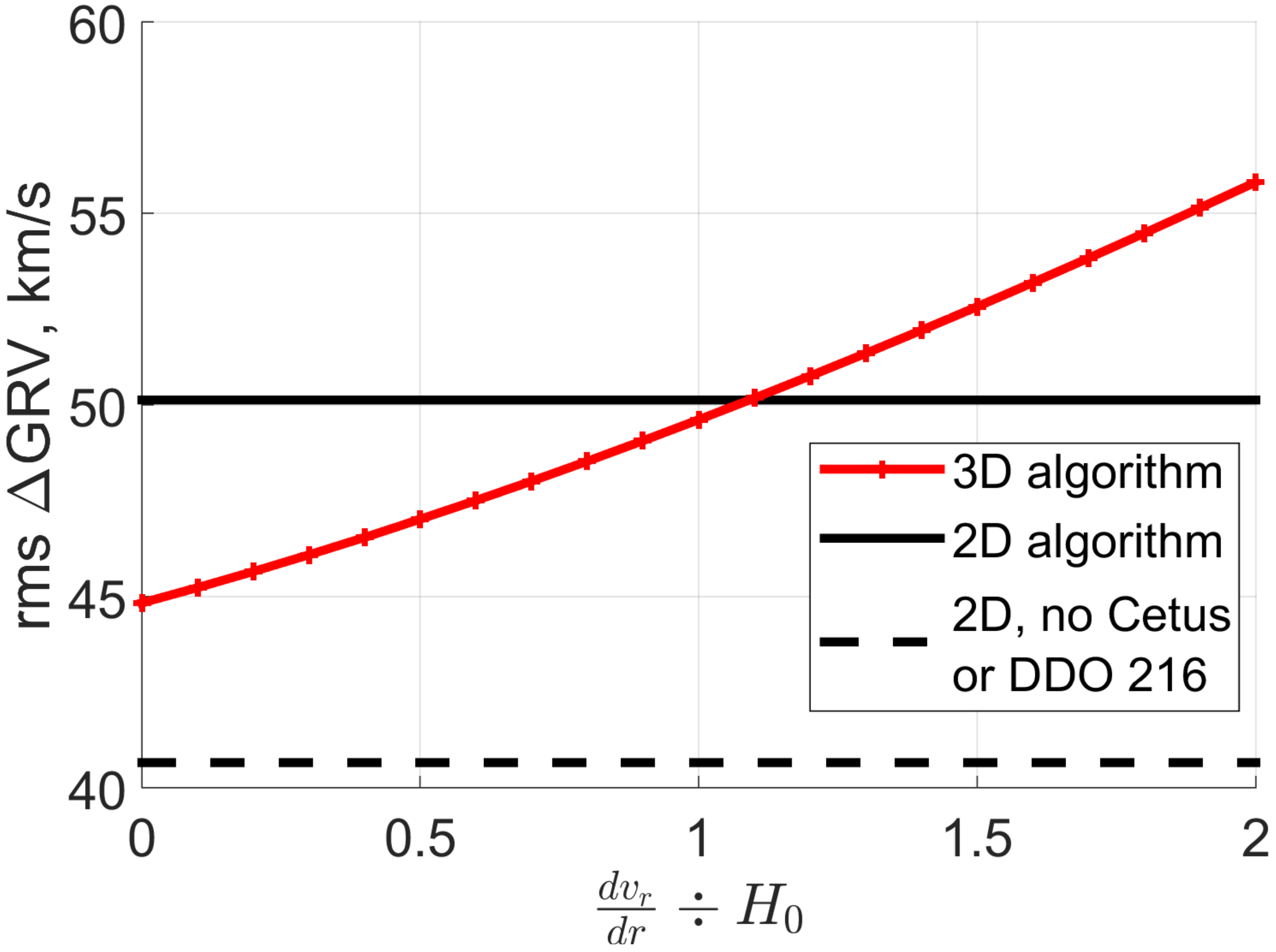}
	\caption{The root mean square value of $\Delta GRV$ is shown here for the best-fitting 2D (black) and 3D (red) models. $\alpha$ governs the way we adjust 3D model predictions to put them on an equal footing with our 2D model (Equation \ref{GRV_adjustment_distance_error}). The adjustment is unnecessary for the latter (see text). This model is likely unreliable close to M31 as it lacks M33, making its predictions for Cetus and DDO 216 unreliable. Results including these galaxies (solid black) and without them (dashed black) are shown. In the 3D model, removing them increases the results by $\ssim 0.7$ km/s, leaving them almost unchanged. This model treats the LG as empty apart from a few point masses. Using a similar assumption in our 2D models would reduce the rms value of $\Delta GRV$ by $\ssim 6$ km/s (not shown).}
	\label{rms_Delta_GRV}
\end{figure}

In \citet{Banik_Zhao_2016}, we added an extra dispersion term to Equation \ref{sigma} and marginalised over other variables to obtain a probability distribution for it. The most likely value we obtained (using the optimal LMC mass) was 40.43 km/s. Using the same target galaxies, the rms dispersion in $\Delta GRV$ with respect to the best-fitting 2D model is 40.65 km/s, almost exactly the same. This suggests that the two statistics are very similar, even though the former uses integration over model parameter space while the latter is based on just one model. Thus, the rms $\Delta GRV$ likely provides a very good guide to the results of a more thorough statistical analysis attempting to pin down how inaccurate our model is as a representation of the data.

After obtaining corrected GRV predictions using Equation \ref{GRV_adjustment_distance_error}, we subtract them from observed GRVs (Equation \ref{GRV_obs}) to obtain a list of $\Delta GRV$s. We determine the rms of these $\Delta GRV$s for a range of plausible assumptions regarding $\alpha$. The precise value does not much affect our overall conclusions (Figure \ref{rms_Delta_GRV}).

For comparison, we also show the result of the same calculation for our best-fitting 2D model using the same target galaxies. This model requires an extremely precise match between their simulated and observed distances, making the result independent of $\alpha$. Although it was technically difficult for us to operate the 3D model in this way, we can gain a conservative lower bound on the rms value of $\Delta GRV$ if we did this by considering the $\alpha = 0$ case in Figure \ref{rms_Delta_GRV}. This corresponds to taking the GRV predictions of the 3D algorithm at face value, even though it has some flexibility with distances. Removing this flexibility can only worsen the agreement between predicted and observed GRVs.

With $\alpha$ irrelevant for our 2D model, the main uncertainty becomes whether Cetus and DDO 216 are included in the analysis as they are very discrepant with this model. We suggest that they should not be included as they are quite close to M31 (Figure \ref{Velocity_field_2D_LMC}). If one also excludes them from the 3D analysis, then the rms value of $\Delta GRV$ for it is hardly affected (it rises $\ssim 0.7$ km/s), thus greatly exceeding the 2D result for the same sample. Even if these galaxies are included, any value of ${\alpha > 1.1}$ implies that the rms $\Delta GRV$ is larger in our 3D analysis.

\begin{table}
	\begin{tabular}{lcc}
	\hline
	Galaxy & $~~\Delta GRV$ & Distance from LG \\
	& $~~$(km/s) & barycentre (Mpc) \\
	\hline
	HIZSS 3 & $~~~123.2 \pm 10.6$ & $1.76 \pm 0.11$ \\
	NGC 3109 & $~~~110.7 \pm 7.3~$ & $1.63 \pm 0.05$ \\
	Sextans A & $~~~~95.1 \pm 7.2~$ & $1.66 \pm 0.02$ \\
	Sextans B & $~~~~75.4 \pm 5.4~$ & $1.71 \pm 0.05$ \\
	Antlia & $~~~~61.6 \pm 8.3~$ & $1.68 \pm 0.06$ \\
	\hline
	UGC 4879 & $~-31.1 \pm 5.5~$ & $1.32 \pm 0.02$ \\
	KKR 3 & $~-33.6 \pm 10.9$ & $2.30 \pm 0.12$ \\
	GR 8 & $~-40.0 \pm 10.5$ & $2.42 \pm 0.12$ \\
	NGC 55 & $~-42.0 \pm 10.4$ & $2.08 \pm 0.11$ \\
	NGC 4163 & $-130.6 \pm 7.7~$ & $2.96 \pm 0.04$ \\
	\hline
	\end{tabular}
	\caption{$\Delta GRV$s with respect to our 3D model for the LG galaxies most discrepant with it (excluding NGC 404 and Leo P due to large distance uncertainties, see text). Errors are estimated using Equation \ref{sigma}. The LG barycentre is put almost exactly at the MW-M31 mid-point (Table \ref{Best_fit_parameters}). Errors in the distance from there are obtained from those on heliocentric distances in the usual way.}
\label{Delta_GRV_3D_list}
\end{table}

The results shown in Figure \ref{rms_Delta_GRV} for the 2D model correspond to a LG filled with matter at the cosmic mean density. To see how much this assumption might affect our results, we previously repeated some of our 2D calculations assuming an empty LG apart from the MW and M31 \citep[][Section 4.1]{Banik_Zhao_2016}. This naturally raises predicted GRVs towards the outskirts of the LG, thereby improving the agreement with observations and reducing the extra dispersion by ${\ssim 6}$ km/s. This reinforces our conclusion that the 3D model does not yield a better match to observations than our previous axisymmetric investigation. In fact, the agreement is slightly worse for the most plausible model assumptions.

Our previous work suggests that we can obtain an error estimate for Figure \ref{rms_Delta_GRV} using the usual rule for the uncertainty in the rms of $N$ independent random variables. In this case, the fractional uncertainty when ${N \gg 1}$ is $\frac{1}{\sqrt{2N}}$, where the number of galaxies is ${N = 34}$. Thus, we expect an uncertainty of $\ssim 6$ km/s, making at least the 3D results inconsistent with the 30 km/s scatter about the Hubble flow found by \citet{Aragon_Calvo_2011}.

In order to estimate uncertainties more rigorously, we run another axisymmetric simulation with parameters chosen to match those in our best-fitting 3D model. This is only possible for some parameters (shown in Table \ref{Best_fit_parameters}) as several relate to particles unique to the 3D model and to motion in 3D.

Using this model, we obtain estimates of how much uncertainty there is on the predicted GRV of each target because of its uncertain position along the line of sight (Equation \ref{sigma_pos}). To obtain the uncertainty on its $\Delta GRV$, we also need to add in quadrature the uncertainty on its observed radial velocity $\sigma_{v_h}$ (Equation \ref{sigma}).\footnote{The actual observational uncertainty is used, not 20 km/s.} We expect this to capture the major observational sources of error.

After determining $\Delta GRV$ and its uncertainty for each target with respect to our 3D model, we can readily see if any galaxies have an unusual redshift for their position. Neglecting a couple of galaxies with large distance uncertainties,\footnote{Leo P and NGC 404 have $\Delta GRV$s of ${72 \pm 32}$ and ${70 \pm 38}$ km/s, respectively} we show the 5 most extreme cases of anomalously high and low GRVs in Table \ref{Delta_GRV_3D_list}. It is apparent that several galaxies have observed GRVs substantially different from that predicted by our best-fitting model. Most of these galaxies have $\Delta GRV > 0$.



We have treated NGC 3109 and Antlia as separate objects. However, they may be gravitationally bound \citep{Van_den_Bergh_1999}. There are indications that they have recently interacted, based on observations of both NGC 3109 \citep{Barnes_2001} and Antlia \citep{Penny_2012}. This is more likely if Antlia is a satellite of NGC 3109. The ${41 \pm 1}$ km/s difference in their GRVs and their $1.19^\circ$ sky separation (corresponding to $\ga 28$ kpc) are likely consistent with this scenario if their heliocentric distances are similar. The distance to Antlia was found to be ${1.31 \pm 0.03}$ Mpc by a study focusing exclusively on this galaxy \citep{Pimbblet_2012}, similar to the ${1.286 \pm 0.015}$ Mpc measured previously by \citet{Dalcanton_2009}. The range of published distances to NGC 3109 is wider than their formal uncertainties, but the most accurate one (based on Cepheid variables) is ${1.300 \pm 0.012}$ Mpc \citep{Aracauria_NGC3109}. Thus, these two galaxies are probably not much more than 40 kpc apart and may well be bound.


Even if it turns out that Antlia should not be treated as an independent object, our overall conclusions should not be much affected because its $\Delta GRV$ is close to the typical $\ssim 50$ km/s (Figure \ref{rms_Delta_GRV}). As a result, the removal of Antlia\footnote{almost 5 magnitudes fainter than NGC 3109 \citep[][Table 3]{McConnachie_2012}} from our sample only reduces the rms value of $\Delta GRV$ by 0.30 km/s. In this case, the appearance of Table \ref{Delta_GRV_3D_list} would also remain similar, with Tucana taking the place of Antlia. Tucana has a $\Delta GRV$ of ${60.3 \pm 7.7}$ km/s and is located ${1.07 \pm 0.05}$ Mpc from the LG barycentre.

We did not put Leo P into Table \ref{Delta_GRV_3D_list} due to a 32 km/s uncertainty on its $\Delta GRV$, almost entirely due to a rather uncertain distance of ${1.72 \pm 0.4}$ Mpc. This is derived from ground-based observations \citep{McQuinn_2013}. However, a more accurate distance measurement has recently been made using the Hubble Space Telescope \citep[${1.62 \pm 0.15}$ Mpc,][]{McQuinn_2015}. Based on how GRV predictions in our axisymmetric model vary with the assumed distance to Leo P, we estimate that this increases its $\Delta GRV$ by ${\ssim 9}$ km/s while reducing the error on it to only ${\ssim 13}$ km/s.

Considering the large difference between simulated and observed distances to Leo P (ruled out at almost ${4\sigma}$ using the newer distance), we looked more closely into how much we should adjust its GRV prediction to make this correspond to its observed position. Normally, we use Equation \ref{GRV_adjustment_distance_error} without worrying too much about the precise value of $\alpha$ as typical distance errors are small. This is not the case here. Assuming that our axisymmetric model (with the same parameters as our 3D model) provides a better guide to how GRV predictions are affected by line of sight distances, it appears that $\alpha$ is overestimated slightly for Leo P. A value of only 1.12 is more appropriate, implying that we reduced its GRV prediction too much to account for it being closer in reality than in our model. As a result, its $\Delta GRV$ is slightly smaller, with a best guess of ${68 \pm 13}$ km/s using the updated distance and method.

Our closer look into Leo P hardly changes our estimate of its $\Delta GRV$ from the ${72 \pm 32}$ km/s assumed in the rest of this work. However, the error budget is more than halved, making it as discrepant with our 3D model as Antlia and Tucana. Importantly, the small change to the result for Leo P and the almost negligible effect of removing Antlia from our sample both lend confidence that our results should not change too much with future improvements to the data and model. This is especially true when one considers that the data for Leo P is particularly inaccurate if using its old distance estimate (Figure \ref{Distance_GRV_correlation}). Almost all other galaxies have substantially smaller observational uncertainties.

\section{Discussion}
\label{Discussion}

Realising the difficulty faced by our axisymmetric model in explaining observations of the LG, we used a 3D model with many times more free parameters (Table \ref{Massive_galaxy_list_3D}). The model was also constrained using more observations, but a lot of these were given quite large uncertainties. For example, we gave all simulated galaxies a HRV error budget of 20 km/s and assumed a 10\% distance uncertainty for galaxies outside the LG (Equation \ref{Distance_uncertainty_perturbers}). We also relaxed the requirement for galaxies to be following the Hubble flow (Equation \ref{Initial_conditions}) at the start of our simulations. It is clear that our best-fitting 3D model has taken full advantage of this liberty (Figure \ref{v_pec_histogram}).

As well as relaxing the constraints on our model in these ways, we used a slightly higher value for the Hubble constant to take advantage of a previous simulation and obtain convergent results faster. We also treated the effect of cosmological acceleration differently (Equation \ref{Equation_of_motion_3D}). In our axisymmetric model, these changes generally increase the predicted GRVs of galaxies near the edge of the LG for fairly intuitive reasons, thus improving the fit to observations \citep[][Sections 4.1 and 4.2]{Banik_Zhao_2016}.

Despite these alterations, our best-fitting 3D model provides a similarly poor match to observations as our best-fitting axisymmetric model (Figure \ref{rms_Delta_GRV}). Some dispersion amongst the $\Delta GRV$s is expected due to observational and modelling uncertainties. Using a rigorous grid investigation of the model parameters, we showed previously that there is an unacceptably large typical mismatch between observations and predictions made by our axisymmetric model \citep[][Section 4]{Banik_Zhao_2016}. Similar conclusions would be obtained considering the best-fitting model alone (Section \ref{3D_results}). We assume that this is also true in our 3D model, justifying our focus on the best-fitting parameter combinations in both models.

The massive galaxies in our models are treated as having a constant mass, implicitly assuming that they formed instantaneously at redshift 9. In reality, galaxies typically gain mass through accretion \citep{Jing_2009}. However, we expect that this should not much affect the present-day velocity of a galaxy required to be at a particular position. This is because the same initial conditions would lead to a different final position, requiring us to adjust the initial conditions so as to counteract this. As a result, the present velocity of a galaxy at a known position is not very sensitive to the forces acting on it at earlier times. Accounting for this `initial condition drag', the effect on the present velocity of a galaxy due to an impulse at some earlier time is $\appropto a^{2.4}$ for an impulse applied when the scale-factor of the Universe was $a$ \citep[][Figure 4]{Banik_Zhao_2016}. To obtain a rough estimate for how much accretion histories might affect our results, we varied the start time of our simulations to correspond to redshift 14 instead of 9 \citep[][Section 4.6]{Banik_Zhao_2016}. Because of the initial condition drag effect, the mismatch between model and observations was only affected by ${\ssim 1}$ km/s.

\begin{figure}
	\centering 
		\includegraphics [width = 8.5cm] {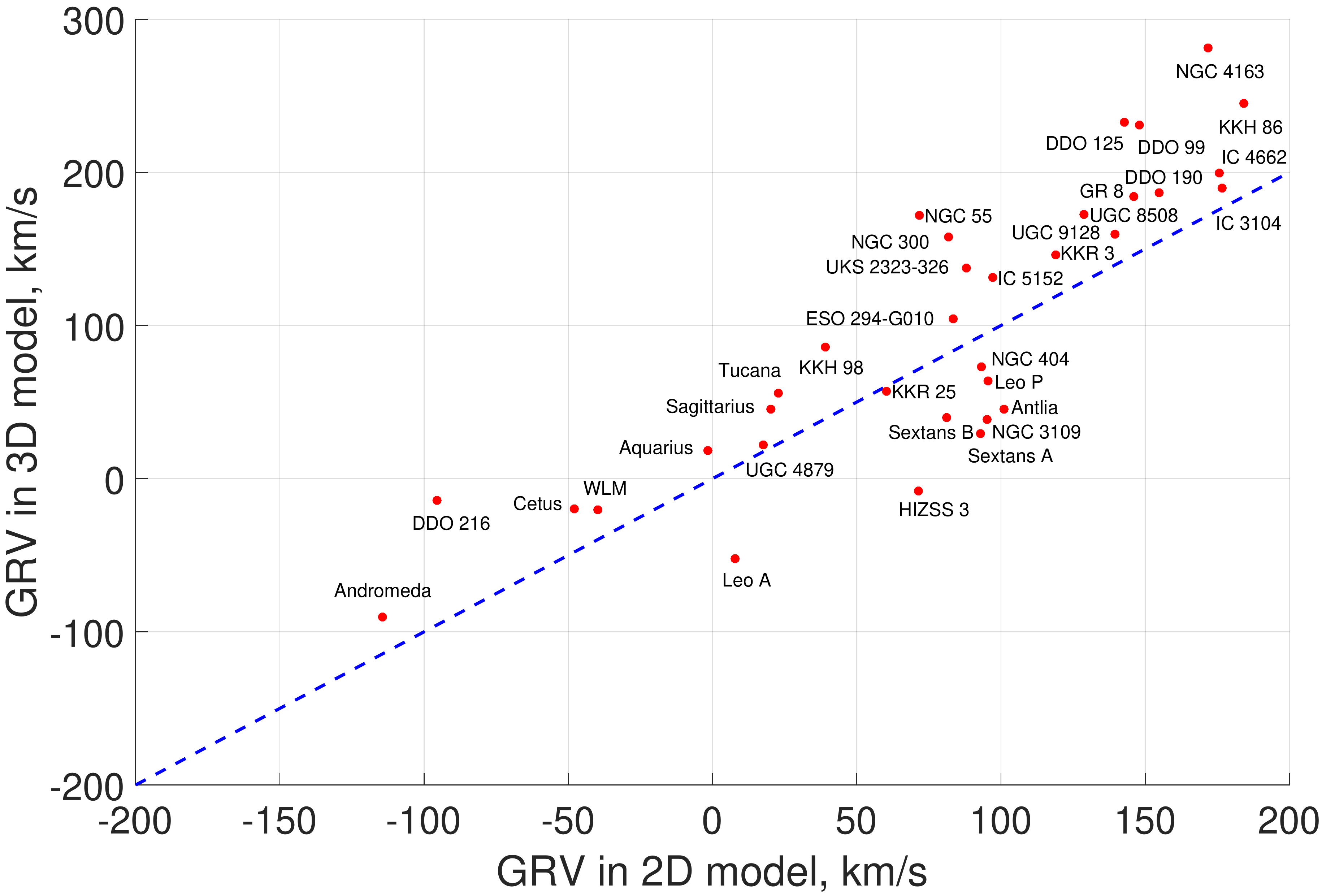}
	\caption{Comparison between the GRVs predicted by our 2D and 3D models with the same model parameters (last column of Table \ref{Best_fit_parameters}). The line of equality is shown in blue. The 2D results may be biased lower towards the edge of the LG because it is assumed to be filled with matter at the same density as the cosmic mean value.}
	\label{Model_comparison}
\end{figure}

Our results must depend somewhat on the assumed mass profile for our simulated massive galaxies. In each case, the profile is constrained observationally by the flatline level of the observed rotation curve of the corresponding galaxy. With this constraint, whatever specific scheme is used for softening the forces close to massive objects (e.g. Equation \ref{Equation_of_motion_3D}), there must be a transition to an inverse square law beyond a distance of ${r_{_S} \ssim \frac{GM}{{v_{_f}}^2}}$. This is only ${\ssim 150}$ kpc for both the MW and M31. Thus, there can be little doubt about the force law further than ${\ssim 1}$ Mpc from them, the region where the galaxies most discrepant with our model seem to lie (Table \ref{Delta_GRV_3D_list}).

Without a grid investigation of the parameters in our 3D model, one might be concerned whether a qualitatively different type of trajectory for some of the massive galaxies might alter the gravitational field in the LG so as to greatly improve the fit to observations. We consider this unlikely for a timing argument analysis such as ours. The reason is that any plausible solution has the MW and M31 turning around just once, while more distant objects outside the LG have not turned around. It is not feasible for the MW and M31 to have undergone a past close flyby in $\Lambda$CDM (or in our model) for any realistic total mass of these galaxies. Thus, the algorithm only needs to solve the continuous problems associated with determining when the turnaround occurred, what the present distance to M31 and its GRV are etc. It does not need to solve discrete problems like how many times the MW and M31 have turned around. This suggests that gradual adjustments to the model parameters should eventually converge on the best-fitting solution, as assumed in the rest of this work. However, it should be borne in mind that the solution presented here is preliminary and a better fit to the observations might eventually be obtained using our model.

Although our 2D and 3D models yield broadly similar conclusions, this is not always the case for individual galaxies. To compare the models in a more direct way, we ran our axisymmetric model using the same parameters as our best-fitting 3D model (Table \ref{Best_fit_parameters}), ignoring features unique to the latter. The GRV predictions of our 3D model were referred to the barycentre of the MW and LMC as the latter is only included indirectly in our 2D model \citep[][Section 4.4]{Banik_Zhao_2016}.

Predictions from both models are compared in Figure \ref{Model_comparison}. The models broadly agree, with an average absolute difference of 44.6 km/s. Interestingly, the 3D results are biased higher by 20.4 km/s. This may be a consequence of the 2D models treating the LG as filled with matter at a density equal to the cosmic mean value. In the 3D model, it is treated as empty apart from a few point masses. It is unclear how much matter is spread diffusely around the LG, but clearly 0 is a lower limit. A larger amount almost certainly worsens the agreement with observations \citep[][Section 4.1]{Banik_Zhao_2016}. This is because the decelerating effect of the extra matter makes it more difficult to explain the high observed GRVs of several LG galaxies. Thus, the 3D results shown in Figure \ref{rms_Delta_GRV} may well underestimate the actual extent of the discrepancy between this model and observations.


To estimate how much this assumption might affect the velocity of a target galaxy currently $d=3$ Mpc away, we determined the time integral of the force on it due to a fixed interior mass $M$ corresponding to the present cosmic mean matter density.
\begin{eqnarray}
	M ~=~ 4\times 10^{12} M_\odot \left( \frac{d}{\text{3 Mpc}} \right)^3
\label{LG_diffuse_mass}
\end{eqnarray}

This is appropriate for a galaxy that is almost following a pure Hubble flow relation at all times, so that its distance in the past is $a\left(t\right)d$. Importantly, the effect of forces acting at earlier times needs to be reduced by a factor of $a^{2.4}$ to account for initial condition drag \citep[][Figure 4]{Banik_Zhao_2016}. Bearing this in mind,
\begin{eqnarray}
	\Delta v \approx \int_{t_{_i}}^{t_{_f}} \frac{GM}{\left(ad\right)^2}\cdot a^{2.4}~dt
\end{eqnarray}

The expansion history of the Universe (Equation \ref{Expansion_history}) can be approximated as
\begin{eqnarray}
	a\left( t \right) ~\approx~ H_{_0}t
	\label{Linear_a_approximation}
\end{eqnarray}

Thus, for simulations starting when ${a \equiv a_{_i}}$, we get that
\begin{eqnarray}
	\Delta v &\approx& \frac{5}{7}\frac{GM}{d^2H_{_0}} \left( 1 - a_{_i}^{1.4} \right) \\
	&=& 19.5 ~\text{km/s}~\text{for }d = 3 \text{ Mpc}
\end{eqnarray}

An effect of this magnitude could go a long way towards explaining why the GRV predictions in our 3D model tend to exceed those in our 2D model. However, there must be other factors as well because a 20 km/s effect due to the different treatment of mass in the LG is only valid for a galaxy currently 3 Mpc away. In reality, most of our targets are at smaller distances (Figure \ref{Velocity_field_2D_LMC}). As a result, they are less affected by a homogeneous distribution of matter because the Shell Theorem implies that the relevant mass ${M \propto d^3}$ (Equation \ref{LG_diffuse_mass}).

One such factor may be that, unlike our 2D model, the 3D model has the freedom to increase predicted GRVs through adjusting the tides raised on the LG by objects outside it (Table \ref{Massive_galaxy_list_3D}). Due to the divergence-free nature of the gravitational field far from these perturbers, one also expects radial velocities to be reduced for LG galaxies in certain directions (Equation \ref{Distant_tide_approximation}). We suppose that our algorithm tries to ensure there are fewer target galaxies $-$ especially those with a high $\Delta GRV$ $-$ towards these directions.


As well as increased flexibility at long range, our 3D model should be more accurate close to the MW and M31 because it directly includes their most massive satellites. This is probably why it achieves a much better fit to the GRV of DDO 216, which is otherwise too high in our 2D model. In our 3D model, it has a close (73 kpc) encounter with M33 almost exactly 8 Gyr ago.

\begin{figure}
	\centering 
		\includegraphics [width = 8.5cm] {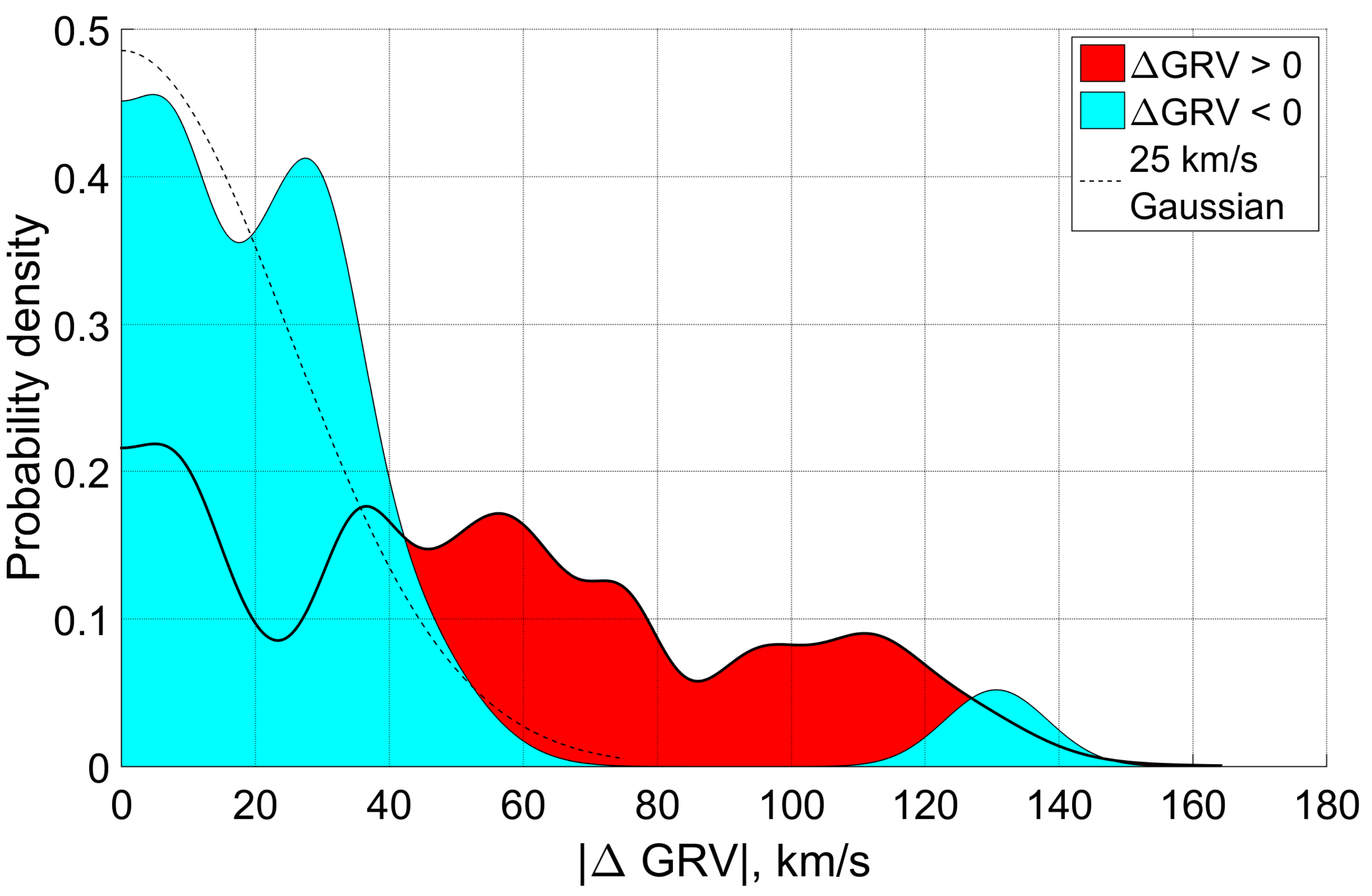}
	\caption{Histogram of $\Delta GRV$s with respect to our 3D model, shown separately according to the sign of $\Delta GRV$. The area of each square corresponds to 2 galaxies. A similar pattern emerges to Figure \ref{Delta_GRV_histogram_LMC_2D}, with the blue bump near 130 km/s caused by NGC 4163. Otherwise, the galaxies with $\Delta GRV < 0$ (solid blue) are well described by a 25 km/s Gaussian (dashed line). This is not the case for galaxies with $\Delta GRV > 0$ (solid red).}
	\label{Delta_GRV_histogram_3D}
\end{figure}

To gain a feel for the overall pattern of discrepancies between our model and observations, we construct a histogram of the $\Delta GRV$s of our target galaxies using a similar procedure to that used for Figure \ref{Delta_GRV_histogram_LMC_2D}. This is shown in Figure \ref{Delta_GRV_histogram_3D}. Because there is some uncertainty in how to convert HRVs into GRVs due to imperfect knowledge of $v_{c, \odot}$, we add 5 km/s in quadrature to the other uncertainties. Although the 3D analysis uses a wider (10 km/s) prior on $v_{c, \odot}$, we choose to stick with the previous value to allow for easier comparison between the histograms.

Both models have a tendency for observed GRVs to exceed predicted ones. The pronounced bump near $\Delta GRV = 80$ km/s in the 2D model gives way to a much broader and shallower bump at a slightly lower $\Delta GRV$ with respect to our 3D model. Neglecting the bump near 160 km/s in the 2D results because it corresponds to DDO 216, the distribution of $\Delta GRV$s extends to even higher values in the 3D model.

Some of the galaxies we identify as having a high $\Delta GRV$ have previously been identified as such. For example, \citet{Pawlowski_McGaugh_2014} identified NGC 3109, Antlia, Sextans A, Sextans B and Tucana as having anomalously high GRVs. With a slightly lower $\Delta GRV$ of ${60.3 \pm 7.7}$ km/s, Tucana is still significantly discrepant with our model.

\citet{Teyssier_2012} identified these 5 objects as possible backsplash galaxies i.e. they have likely passed within the virial radius of the MW or M31. There, they may have passed close to a massive satellite such as M33. Despite allowing for such trajectories, we are still unable to provide a good match to the observed GRVs of several LG dwarf galaxies. We also note that \citet{Shaya_2011} found it very difficult to incorporate Sextans A and B into their dynamical model of the LG, suggesting that it could not easily explain their motions. Thus, the high velocity galaxy problem appears to persist even with a 3D model.

Interestingly, some of these galaxies appear to be correlated in phase space. In particular, the galaxies in the NGC 3109 association (NGC 3109, Antlia, Sextans A, Sextans B and Leo P) seem to lie very close to a line \citep{Bellazzini_2013}. Their radial velocities also closely follow a tight trend of increasing for galaxies further from the LG. We hope to investigate this further in a future publication.

The most extreme outlier with respect to our model is actually a galaxy with negative $\Delta GRV$. This is NGC 4163, a galaxy which appears unusual even in the almost model-independent analysis shown in Table \ref{Two_stream_investigation_results}. Here, it is apparent that two nearby galaxies have a much higher $\Delta GRV$, suggesting that GRVs of galaxies near NGC 4163 are higher than its GRV. This is also apparent when considering the radial velocities of other galaxies at similar heliocentric distances in the Canes Venatici I cloud \citep[][Table 2]{Makarov_2013}. Thus, leaving aside any models, it appears that the HRV of NGC 4163 is $\ssim 100$ km/s less than that of neighbouring galaxies.

One possible explanation is that NGC 4163 was flung towards the LG as a result of a gravitational slingshot interaction in a nearby galaxy group. It is also possible that NGC 4163 had a more recent interaction with a galaxy close to it, as perhaps suggested by its recent starburst \citep{McQuinn_2009}. In fact, our 3D model has DDO 99 passing within 50 kpc of it, although this happened in ancient times (when ${a = 0.32}$). Nonetheless, it is possible that our model has got the timing of this encounter wrong, especially as it does not simulate the effects of any such encounter because both galaxies are treated as test particles. Certainly such a close encounter is rare in our models. If it did happen, then the fact that NGC 4163 is almost a magnitude fainter \citep[][Table 3]{McConnachie_2012} suggests that its dynamics would be affected to a greater extent. Thus, several explanations are possible for its anomalously low GRV.

Interactions amongst galaxies in a neighbouring group can fling a galaxy towards the LG, leading to it having an unusually low GRV. However, it is very difficult to explain a galaxy with anomalously high GRV in this way. This is possible only if the galaxy has crossed the LG and is now heading away from it again. It is not feasible for a galaxy like NGC 3109 to cover such a large distance in the time since the Big Bang. To see why, consider that it is 1.6 Mpc from the LG, such that the Hubble velocity is $\ssim 110$ km/s. Given an extra radial velocity of 120 km/s, we see that the galaxy can have covered perhaps twice as much distance as a typical galaxy on the Hubble flow.\footnote{A more accurate estimate is given in Equation \ref{Comoving_displacement_estimate}.} This means that the anomalous motion of NGC 3109 could not have originated much further from the LG than its present distance, although it could have done so on the opposite side to its current location.

Furthermore, even if this scenario was plausible, it is clear that it would be rarer than a situation where a galaxy is flung towards the LG and we observe it on the way in. Thus, we might see several galaxies with anomalously high GRV, but we would expect to see even more with anomalously low GRV. The opposite seems to be the case (Figure \ref{Delta_GRV_histogram_3D}).

\begin{figure}
	\centering 
		\includegraphics [width = 8.5cm] {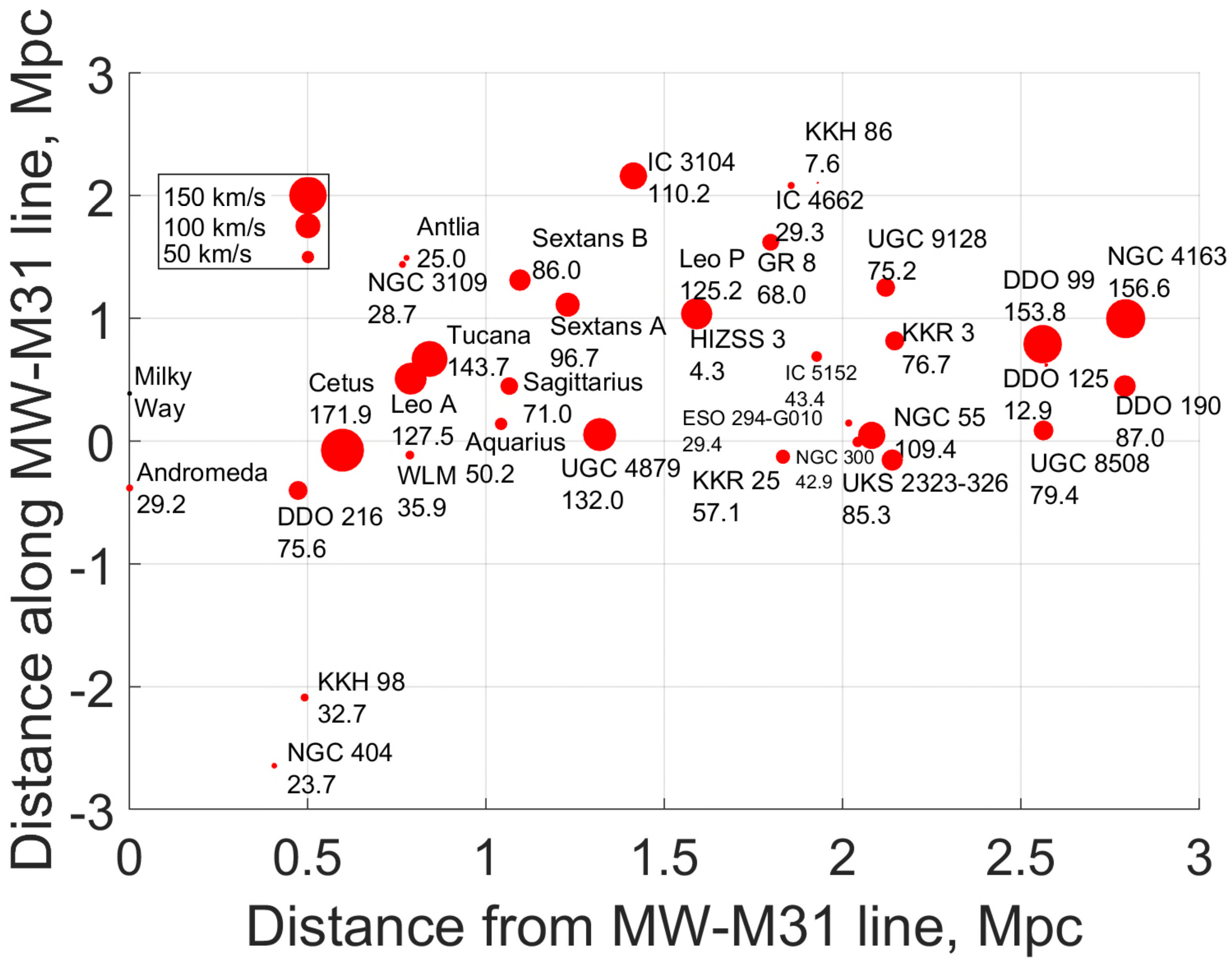}
	\caption{Velocities of our target galaxies out of the plane they define with the MW and M31 are shown here using marker sizes. This information is also given as a number (km/s) below the corresponding galaxy name. Galaxy positions are shown relative to the MW$-$M31 line.}
	\label{Non_axisymmetry}
\end{figure}

Using a 3D model, we can quantify how fast target galaxies move out of the plane they define with the MW and M31. Referring velocities to the barycentre of the MW and LMC, we obtain the results shown in Figure \ref{Non_axisymmetry}. Part of the reason for motions deviating from axisymmetry with respect to the MW-M31 line is that the direction of this line has rotated slightly because the MW and M31 are not on a purely radial orbit. However, this is only a $\ssim 30$ km/s effect \citep{Van_der_Marel_2012}. Thus, the explanation must lie mostly with the non-axisymmetric gravitational field caused by massive objects far from the MW$-$M31 line. Massive satellite galaxies such as the LMC and M33 must also play a role.

The substantial non-axisymmetric motions suggested by our analysis are probably required in order to boost the centrifugal force, helping explain the high GRVs of several LG galaxies. Some of this tangential motion may be a relic of peculiar velocities at the start of our simulation (Figure \ref{v_pec_histogram}), especially when considering that Hubble drag does not dissipate specific angular momentum \citep[Equation 10 of][]{Banik_Zhao_2016}. Although the expansion of the Universe naturally reduces tangential speeds, this is not as important an effect as one might think. For example, our best-fitting 3D model has NGC 3109 starting off 593 kpc from the MW and being 1.37 Mpc away currently. This means that its distance has increased ${<2.5\times}$ despite the Universe overall having expanded by a factor of 10.

In the future, such non-axisymmetric motions should become detectable based on proper motion measurements. A good target for this may be DDO 216 due to its relative proximity. Our model implies that it had a past close encounter with M33, constraining the possible trajectories of these two galaxies and also indirectly forcing some limits on the past position of M31.

The larger expected velocities of Cetus and Tucana out of the plane they define with the MW and M31 should make their proper motions similar in magnitude to that of DDO 216, despite a greater distance. Tucana is one of the nearest galaxies with an unusual GRV in our analysis. But several galaxies with much larger $\Delta GRV$ lie just a little further away, in the vicinity of NGC 3109 (Figure \ref{Distance_GRV_correlation}). Its proper motion promises to be an extremely interesting constraint on any models of the LG \citep{Pawlowski_McGaugh_2014}.


\subsection{Effects of the Great Attractor and the Virgo Cluster}
\label{Great_Attractor}

A 3D model including several objects beyond the LG allows for a much more rigorous handling of the tides they raise within it. Such effects tend to be larger at greater distances. Thus, it is interesting that the most discrepant objects found by our previous axisymmetric analysis tend to lie towards the edge of the LG \citep[][Figure 10]{Banik_Zhao_2016}. A similar trend is apparent when the LMC is included indirectly (Figure \ref{GRV_LCDM_comparison_Cetus}), bearing in mind that a high GRV generally implies a greater distance and that the absence of M33 from the model makes it less reliable for galaxies close to M31 (e.g. Cetus and DDO 216).

In our 3D model, despite handling tides more rigorously, we found a similarly poor match between observed and model-predicted GRVs (Figure \ref{rms_Delta_GRV}). However, the model does not include all the mass concentrations outside the LG which may be relevant to our analysis. An important example is the Great Attractor \citep[GA,][]{Mieske_2005}. This is thought to be primarily responsible for the $\ssim 630$ km/s magnitude of $\bm{v}_{pec, LG}$, the velocity of the LG as a whole with respect to the surface of last scattering \citep{Kogut_1993}.

To estimate the effect of the GA on GRVs of objects within the LG, we use the distant tide approximation. Treating LG galaxies as freely falling in the gravitational field of a distant point mass, the change in the GRV of a target galaxy due to the GA is given by
\begin{eqnarray}
	\Delta GRV_{GA} = \left(3 \cos^2 \theta - 1 \right) \beta_{_{IC}} \frac{v_{pec, LG} ~d}{d_{GA}} ~~~\text{for } d \ll d_{GA}
	\label{Distant_tide_approximation}
\end{eqnarray}

$d_{GA}$ is the distance to the GA while $\theta$ is the angle on our sky between it and the target galaxy, which is at a heliocentric distance $d$. The parameter ${\beta_{_{IC}} \approx 0.76}$ accounts for initial condition drag.

We assume the GA caused the LG to gain a peculiar velocity of $v_{pec, LG} = 630$ km/s. The direction towards the GA is taken to be $l = 325^\circ$, $b = -7^\circ$ in Galactic co-ordinates \citep{Kraan_Korteweg_2000}. This is similar to the direction of $\bm{v}_{pec, LG}$. The discrepancy is likely caused by less massive objects closer to the LG. For example, Centaurus A, M81, IC 342 and the Virgo Cluster are all in the northern Galactic hemisphere, partly explaining why $\bm{v}_{pec, LG}$ points further north than the direction towards the GA.


To estimate $\beta_{_{IC}}$, we construct a basic simulation involving a particle of mass $M$ in an otherwise homogeneous Universe. A test particle some distance $r$ away is required to satisfy $r = d$ with a peculiar velocity close to 630 km/s at the present time. This is achieved by varying its position when the cosmic scale-factor ${a = 0.1}$, at which time the particle satisfied Equation \ref{Initial_conditions}. In Section 2.1 of \citet{Banik_Zhao_2016}, we used General Relativity to show that the test particle satisfies
\begin{eqnarray}
	\ddot{r} ~~=~~ \frac{\ddot{a}}{a}r ~-~ \frac{GM}{r^2}
\end{eqnarray}

We solve this problem using a Newton-Raphson procedure targeting a present distance of $d = 84$ Mpc. We then repeat the calculation with a slightly different $d$. Without the GA, the relative radial velocity of the particles would be $H_{_0}r$. Thus, we determine the effect of the GA using
\begin{eqnarray}
	\Delta GRV_{GA} &=& \Delta v_{pec}~~\text{ where} \\
	v_{pec} &\equiv& \dot{r} - H_{_0}r
    \label{v_pec}
\end{eqnarray} 

In this way, we find that $\beta_{_{IC}} \approx 0.76$ for both cosmological models given in Table \ref{Best_fit_parameters}. Assuming $\Delta GRV_{GA}$ has the same angular dependence as the tidal field that causes it, we suppose that this calculation is sufficient to determine the result for all angles $\theta$. This should be valid as long as the tides raised by the GA can be approximated as linear in position, which is reasonable within the LG.

For $\theta$ close to 0 or $180^\circ$, the GA tends to increase GRVs. However, for $\theta$ close to $90^\circ$, the GA reduces GRVs. This is because both the MW and the target galaxy accelerate towards the perturber at similar rates. As their co-moving distance from the GA decreases, so also does their co-moving distance from each other. Thus, the GA can only increase GRVs along one direction. It must reduce them along the other two, albeit by half as much. This is due to the divergence-free nature of the gravitational field far from its source. As a result, an external tidal field on the LG does not readily resolve the high velocity galaxy problem within it as these galaxies lie in several quite different sky directions (Table \ref{Delta_GRV_3D_list}).

\begin{figure}
	\centering 
		\includegraphics [width = 8.5cm] {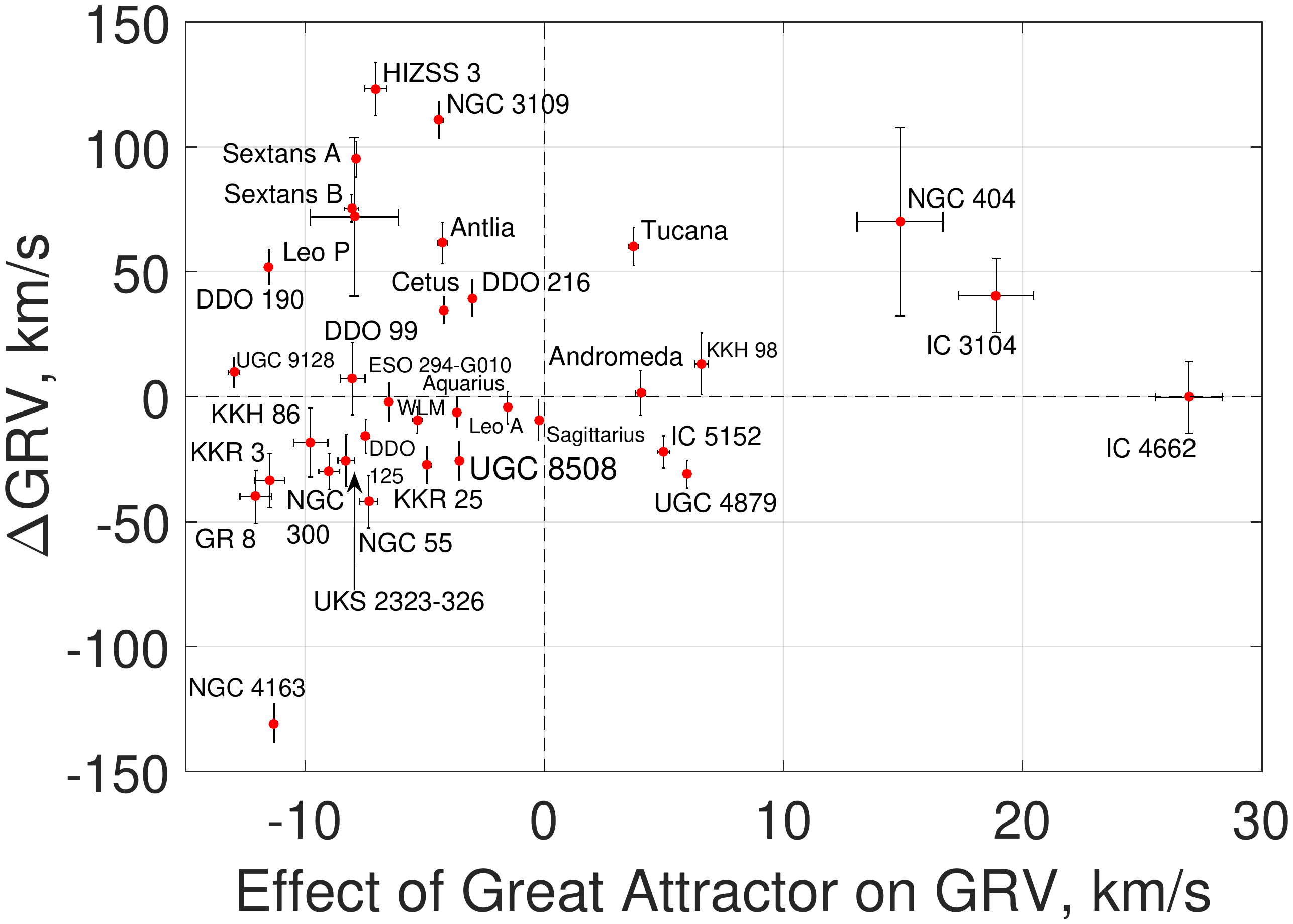}
	\caption{$\Delta GRV$s are shown against our estimate for how much the Great Attractor might have increased the GRV of each galaxy (Equation \ref{Distant_tide_approximation}). Errors are correlated because a larger distance increases the effect of the GA while reducing $\Delta GRV$.}
	\label{Great_Attractor_Peebles_Effect_76}
\end{figure}

Nonetheless, one can hope that the GA helps with some of the most problematic cases. To see if this is likely, we compare the effect of the GA predicted by Equation \ref{Distant_tide_approximation} and the $\Delta GRV$ of each galaxy. This is shown in Figure \ref{Great_Attractor_Peebles_Effect_76}. Here, we have propagated distance uncertainties in the usual way. The error budget on each $\Delta GRV$ accounts for uncertainties in the corresponding HRV and distance measurements (Equation \ref{sigma}). A different GA distance to the assumed 84 Mpc causes a rescaling of its effects.

\begin{figure}
	\centering 
		\includegraphics [width = 8.5cm] {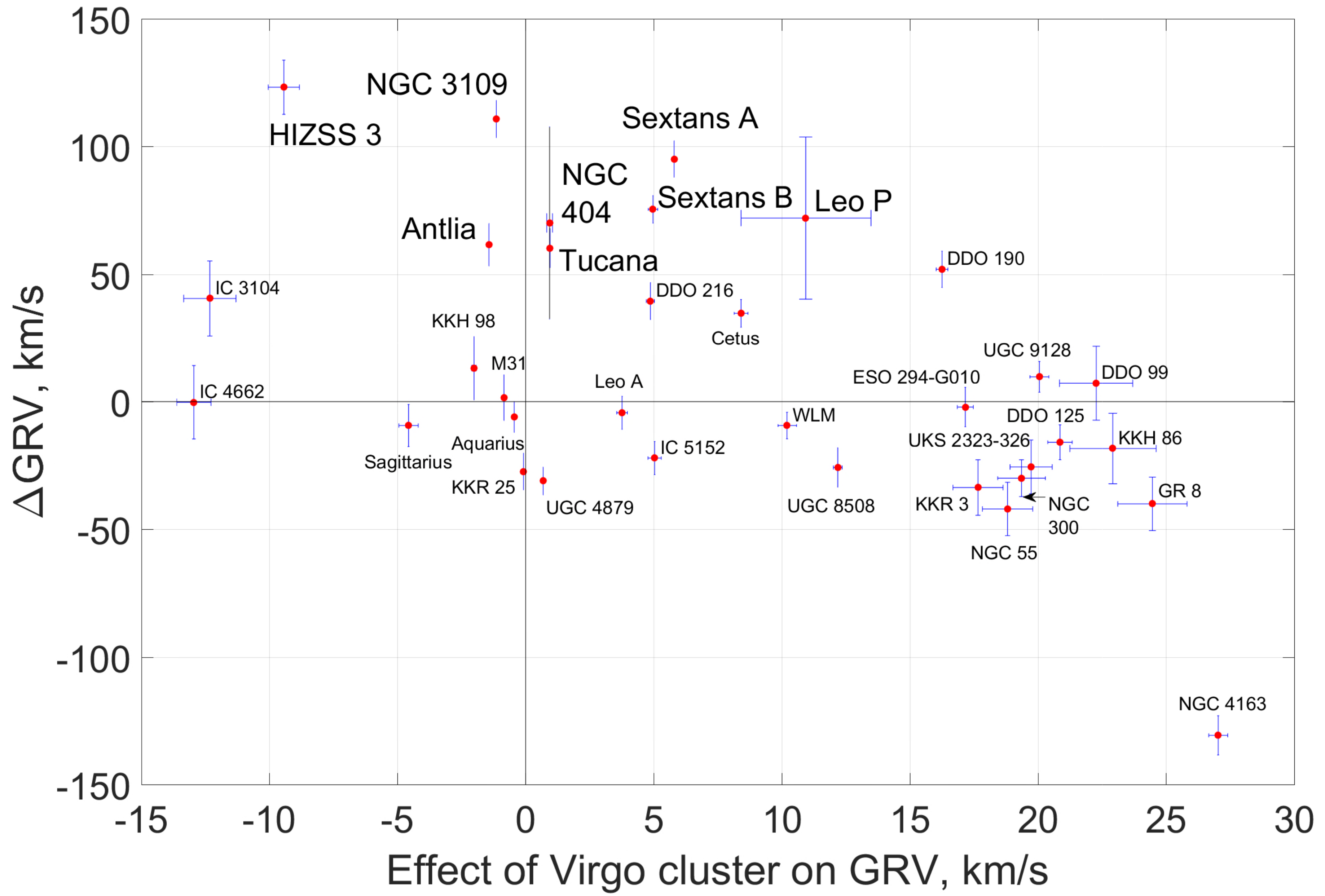}
	\caption{Similar to Figure \ref{Great_Attractor_Peebles_Effect_76}, for the Virgo Cluster. We use the same tidal field strength of 5.7 km/s/Mpc for target galaxies that make a right angle at the MW with the direction towards the Virgo cluster \citep[$l = 283.8^\circ$, $b = 74.4^\circ$,][]{Nezri_2012}. Galaxies with a high $\Delta GRV$ have been emphasized.}
	\label{Virgo_Effect_Peebles_76}
\end{figure}


Equation \ref{Distant_tide_approximation} suggests that the GA should actually have \emph{reduced} the GRV of most or all of the galaxies with the highest $\Delta GRV$ (listed in Table \ref{Delta_GRV_3D_list}). This remains the case if Antlia is treated as a low mass satellite of NGC 3109 and excluded from our sample \citep{Van_den_Bergh_1999}. Thus, far from helping to resolve the high velocity galaxy problem, consideration of the GA appears to make it worse.


Only one galaxy with $\Delta GRV < 0$ has a GRV so low as to be problematic given the expected $\sim$25 km/s accuracy of our model \citep[][Table 3]{Banik_Zhao_2016}. This is NGC 4163, which we have discussed previously. Our results suggest that its GRV would be reduced by the GA, thus helping to explain its low observed GRV. However, the effect is almost an order of magnitude too small.

It is straightforward to consider all the LG target galaxies in our sample, not just those substantially discrepant with our best-fitting model. After adjusting GRV predictions for the GA using Equation \ref{Distant_tide_approximation} with different assumptions regarding the parameter $\alpha$ (Equation \ref{GRV_adjustment_distance_error}), we found that the rms $\Delta GRV$ increased very slightly.
\begin{eqnarray}
	\text{Increase in rms}~\Delta GRV ~\approx~\left(0.84 - 0.36\alpha \right) ~\text{km/s}
	\label{GA_final_result}
\end{eqnarray}

The very small effect of the GA on the rms $\Delta GRV$ could well be a sign that our algorithm has utilised its many degrees of freedom to adjust the tidal field acting on the LG so as to best match observations.

Taken at face value, our analysis suggests that even galaxies at the edge of the LG $\ssim 3$ Mpc away would only have $\Delta GRV_{GA} \sim 35$ km/s at most. Typical effects would be $\ssim \frac{1}{4}$ as much due to smaller distances and projection effects. Considering that several galaxies have a much larger $\Delta GRV$ such that even its rms value exceeds 40 km/s, the effect of the GA is too little to substantially alter our conclusions.

However, cosmological simulations based on $\Lambda$CDM suggest that structure tends to form along filaments \citep[e.g.][Figure 1]{Springel_2005}. Thus, there may well be additional structures along the line of sight towards the GA (or on the opposite side) at smaller distances. The effect of these structures on the LG would have a similar angular dependence as that of the GA, but possibly with an increased magnitude due to the smaller distance. This suggests that we may have underestimated the effects shown in Figure \ref{Great_Attractor_Peebles_Effect_76}.

Besides the GA, another structure outside our analysis which may be important to the LG is the Virgo Cluster (VC). Using the same tidal field strength as for the GA, we estimated how much the VC could affect GRVs of objects in the LG (Figure \ref{Virgo_Effect_Peebles_76}). In this case, the magnitude of the effect is not known so well, making it more useful to focus on its sign as this depends only on the sky positions of the relevant objects. It is apparent that the VC may have some success at reducing the $\Delta GRV$s of Sextans A, B and Leo P but would not much help with the other high $\Delta GRV$ galaxies. The overall trend is less promising as most galaxies are in the `wrong' quadrants $-$ either they have an anomalously high GRV but the VC would reduce their GRV, or vice versa. This is not the case with the GA, which would likely reduce GRVs of most LG galaxies (Figure \ref{Great_Attractor_Peebles_Effect_76}) and thus have little overall effect on how well our model matches observations (Equation \ref{GA_final_result}).


So far, we assumed that the GA and VC have always been in the same direction. Due to their large distances, they are almost certainly very close to the Hubble flow today. This was likely the case over the vast majority of the age of the Universe. It may not have been so at early times. However, because of initial condition drag, gravitational forces then have only a very small effect on present peculiar velocities \citep[e.g.][Figure 4]{Banik_Zhao_2016}. Thus, the crucial ingredient in predicting how the LG should be affected by e.g. the GA is its present position.


\subsection{Modified Gravity}
\label{MOND}

The high GRVs of some LG galaxies must have been caused by forces acting on them which our model does not account for. The nature of these forces might be better understood if we had an idea of the spacetime location where they acted. Thus, it is necessary to estimate where these galaxies may have been. For the sake of clarity, we will focus on NGC 3109 because it has a high $\Delta GRV$ in our model (Table \ref{Delta_GRV_3D_list}) and was also identified as having a rather high GRV in previous works \citep{Pawlowski_McGaugh_2014, Teyssier_2012}.

In this regard, it is helpful to define co-moving positions $\bm{x}$ which do not change for particles in a homogeneous Universe. Thus, they are related to physical positions $\bm{r}$ by
\begin{eqnarray}
	\bm{x} ~\equiv~ \frac{\bm{r}}{a \left( t \right)}
	\label{Co_moving_position}
\end{eqnarray}

As we are dealing with unknown forces, we only seek a rough estimate of the co-moving displacement $d$ of NGC 3109. We assume it is so distant as to be following the Hubble flow in our model, such that ${d \approx 0}$ in it. The actual value of $d$ is estimated based on integrating the trajectory of NGC 3109 backwards in time, with the effect of gravity crudely included by using its $\Delta GRV$ instead of its observed peculiar velocity as a present boundary condition. For simplicity, we only consider motion along the line between the LG barycentre and the present position of NGC 3109. Thus, its present peculiar velocity (Equation \ref{v_pec}) for the purposes of this section is estimated as 
\begin{eqnarray}
	v_{pec,0} ~\approx~ \Delta GRV
\end{eqnarray}

Due to the effect of Hubble drag \citep[e.g.][Equation 24]{Banik_Zhao_2016}, the peculiar velocity of a free particle changes with time according to $v_{pec} \propto a^{-1}$. Furthermore, even the same peculiar velocity corresponds to a more rapidly changing co-moving position in the past (Equation \ref{Co_moving_position}). Thus, integrating between some past time ${t_{_i}}$ and the present time ${t_{_0}}$, we estimate that the co-moving displacement
\begin{eqnarray}
	d &=& \int_{t_{_i}}^{t_{_0}} \frac{v_{pec}}{a}~dt\\
	&=& \int_{t_{_i}}^{t_{_0}} \frac{v_{pec,0}}{a^2}~dt
	\label{Comoving_displacement}
\end{eqnarray}

For a rough estimate, it is acceptable to approximate the expansion history of the Universe as linear in time (Equation \ref{Linear_a_approximation}). During the time period of interest (${a \ga 0.2}$), the results obtained in this way for ${\int_{t_{_i}}^{t_{_0}} a^{-2}~dt}$ are accurate to within $\ssim 10\%$. The error is much smaller if $a_{_i} \equiv a \left( t_{_i} \right)$ is close to 0.2 or 1. This allows us to solve Equation \ref{Comoving_displacement}.
\begin{eqnarray}
	\label{Comoving_displacement_estimate}
	d ~&=&~ \frac{v_{pec,0}}{H_{_0}}\left( {a_{_i}}^{-1} ~-~ 1 \right) \\
	&=&~\frac{v_{pec,0}z}{H_{_0}} ~~\text{ where} \\
	z ~&\equiv &~ {a_{_i}}^{-1} ~-~ 1
\end{eqnarray}


To simplify our discussion, we assume that the high $\Delta GRV$ of NGC 3109 was caused by forces acting over a small fraction of the Hubble time, e.g. due to an encounter with a massive object. There is a trade-off between how long ago these forces acted and their total impulse. M31 is the fastest rotating LG galaxy, with a circular velocity of $\ssim 225$ km/s \citep{Carignan_2006}. An impulse of twice this is only possible for an object not eventually accreted if it is on a circular orbit. Thus, we suppose that the impulse could not feasibly have exceeded triple the $\ssim 110$ km/s $\Delta GRV$ of NGC 3109, implying ${a_{_i} > \frac{1}{3}}$. In the limiting case, an unexplained impulse of ${\ssim 330}$ km/s would have to be acquired when the redshift ${z = 2}$, presumably in a gravitational interaction.

Given that 110 km/s is very close to the Hubble flow rate at the distance of NGC 3109, we can set $v_{pec,0} \approx H_{_0}d_{_0}$, where the present co-moving/physical distance from the LG barycentre to NGC 3109 is $d_{_0}$. Thus, its co-moving displacement $d$ could not have much exceeded twice this, in which case the unexplained impulse occurred $\la 1.7$ co-moving Mpc from the LG, possibly on the opposite side to where NGC 3109 currently lies. This makes it very difficult to understand how the present motion of NGC 3109 came about if one looks for an explanation outside the LG. Furthermore, the effect of gravity implies that galaxy groups outside it must have started at a slightly larger co-moving separation with it than they presently have. For example, our model indicates that the co-moving distance between the MW-M31 mid-point\footnote{a reasonable estimate for their barycentre as our model prefers them to have nearly equal masses (Table \ref{Best_fit_parameters})} and Cen A has decreased from 4.98 Mpc to 4.04 Mpc since redshift 9.

A more plausible scenario might be that the MW and/or M31 are the massive object(s) responsible for the anomalous kinematics of NGC 3109. In this case, the missing ingredient is an impulse close to the LG barycentre, since which time the co-moving displacement ${d \approx d_{_0}}$. For this to occur, we need to set $a_{_i} = \frac{1}{2}$ in Equation \ref{Comoving_displacement_estimate}, corresponding to $\ssim 8$ Gyr ago. The conclusion that NGC 3109 was likely close to the LG barycentre at this time has previously been reached using simpler methods \citep{Pawlowski_McGaugh_2014}.

However, it would be very unusual if the major LG galaxies were responsible for the high $\Delta GRV$ of NGC 3109. After all, our model directly includes the MW and M31 as well as their most massive satellites (Table \ref{Massive_galaxy_list_3D}). Nonetheless, gravitational slingshot interactions with these objects could well lead to high GRVs, as occurs close to the LG barycentre (bottom panel of Figure \ref{Velocity_field_2D_LMC}). Thus, increasing the efficiency of this process might help to explain the observations. The energy gained in such interactions is reliant on the gravitational potential of the massive body being time-dependent due to its motion. This suggests that the relative motion of the MW and M31 might have been much faster in the past than implied by our model. Given their known relative velocity at present, this implies a rather high mutual acceleration. Therefore, we need to consider whether we have correctly understood the gravitational effect of the MW and M31 on each other. 

So far, our discussion has been restricted to models based on $\Lambda$CDM. We have seen that it faces difficulties in explaining the dynamics of LG galaxies, both using a 3D model and a thorough grid investigation of the parameters in an axisymmetric model \citep{Banik_Zhao_2016}.

Beyond the LG, some remarkably tight correlations exist between the dynamics of galaxies and the distribution of their luminous matter \citep[e.g.][and references therein]{Famaey_McGaugh_2012}. This has recently been confirmed and further tightened based on near-infrared observations by the Spitzer Space Telescope \citep{McGaugh_2016}. These correlations were unexpected in the context of $\Lambda$CDM. However, many of the trends were predicted a priori using Modified Newtonian Dynamics \citep[MOND,][]{Milgrom_1983}. Thus, we consider whether this theory may shed light on the high velocity galaxy problem. Our reasoning will be similar to that in Section 4.6 of \citet{Banik_Zhao_2016}.

MOND imposes an acceleration-dependent modification to the usual Poisson Equation of Newtonian gravity \citep{Bekenstein_Milgrom_1984, QUMOND}. In spherical symmetry, the result is that the gravitational field $g$ at distance $r$ from an isolated point mass $M$ transitions from the usual inverse square law at short range to
\begin{eqnarray}
	g ~=~ \frac{\sqrt{GMa_{_0}}}{r} ~~~~~\text{for } ~r \gg \frac{\sqrt{GM}}{a_{_0}}
	\label{Deep_MOND_limit}
\end{eqnarray}

Here, $a_{_0}$ is a fundamental acceleration scale of nature. Empirically, $a_{_0} \approx 1.2 \times {10}^{-10} $ m/s$^2$ to match galaxy rotation curves \citep{McGaugh_2011}. At this value, there is a remarkable coincidence with the acceleration at which the energy density in a classical gravitational field becomes comparable to the dark energy density $u_{_\Lambda}$.\footnote{Dark energy is required to explain why $\ddot{a}>0$ despite the attractive effect of gravity \citep{Riess_1998}.} Thus,
\begin{eqnarray}
	\frac{g^2}{8\rm{\pi}G} ~<~ u_{_\Lambda}c^2 ~~\Leftrightarrow~~ g ~\la~ 2\rm{\pi}a_{_0}
\end{eqnarray}

This suggests that MOND may be caused by quantum gravity effects \citep[e.g.][]{Milgrom_1999, Pazy_2013, Verlinde_2016}. Regardless of the underlying microphysical explanation, at sufficiently low acceleration, MOND gravity from a point mass follows Equation \ref{Deep_MOND_limit} as long as gravity from other objects is negligible.

The external gravitational field on the LG can be estimated based on its peculiar velocity of $\ssim 630$ km/s relative to the surface of last scattering \citep{Planck_2013}. As might be expected, this shows the LG to be fairly isolated \citep{Famaey_2007}. Thus, the force between the MW and M31 declines much slower with their separation than in $\Lambda$CDM, especially in the range between their virial radii ($\ssim 150$ kpc in our models) and their actual separation \citep[${783 \pm 25}$ kpc,][]{McConnachie_2012}.\footnote{At still greater distances, the force would transition to the usual inverse square law, but with a much higher normalisation than in Newtonian gravity \citep{Milgrom_1986}.}

If correct, the much stronger force between these galaxies has dramatic consequences for the whole LG. This is because the MW$-$M31 orbit is almost radial \citep{Van_der_Marel_2012}. As a result, MOND implies a past close flyby encounter between them ${9 \pm 2}$ Gyr ago \citep{Zhao_2013}. The tidal tails expelled from the disks of these galaxies during their interaction may be responsible for the thin co-rotating system of satellites around the MW \citep[e.g.][]{Kroupa_2013} and the similar system around M31 \citep{Ibata_2013}. This interaction may also have formed the thick disk of the MW \citep{Gilmore_1983}, a structure which seems to have formed fairly rapidly from the thin disk ${9 \pm 1}$ Gyr ago \citep{Quillen_2001}. More recent investigations also suggest a fairly rapid formation timescale \citep{Hayden_2015}. The disk heating which likely formed the thick disk appears to have been stronger in the outer parts of the MW, characteristic of a tidal effect \citep{Banik_2014}. This may be why it has a larger scale length than the thin disk of the MW \citep{Juric_2008, Jayaraman_2013}.

At the point of closest approach, the relative velocity of the MW and M31 would have been $\ssim 600$ km/s \citep{Zhao_2013}. Such fast motions could lead to very powerful gravitational slingshot encounters. The limiting factor might even have been their circular rotation velocities rather than the motions of their centres of mass. If we assume a maximum impulse of ${\ssim v_{_{f,M31}}}$ and an encounter when $a_{_i} = \frac{1}{2}$, then it is easy to see why there are no galaxies with $\Delta GRV \ga 120$ km/s.

\begin{figure}
	\centering 
		\includegraphics [width = 8.5cm] {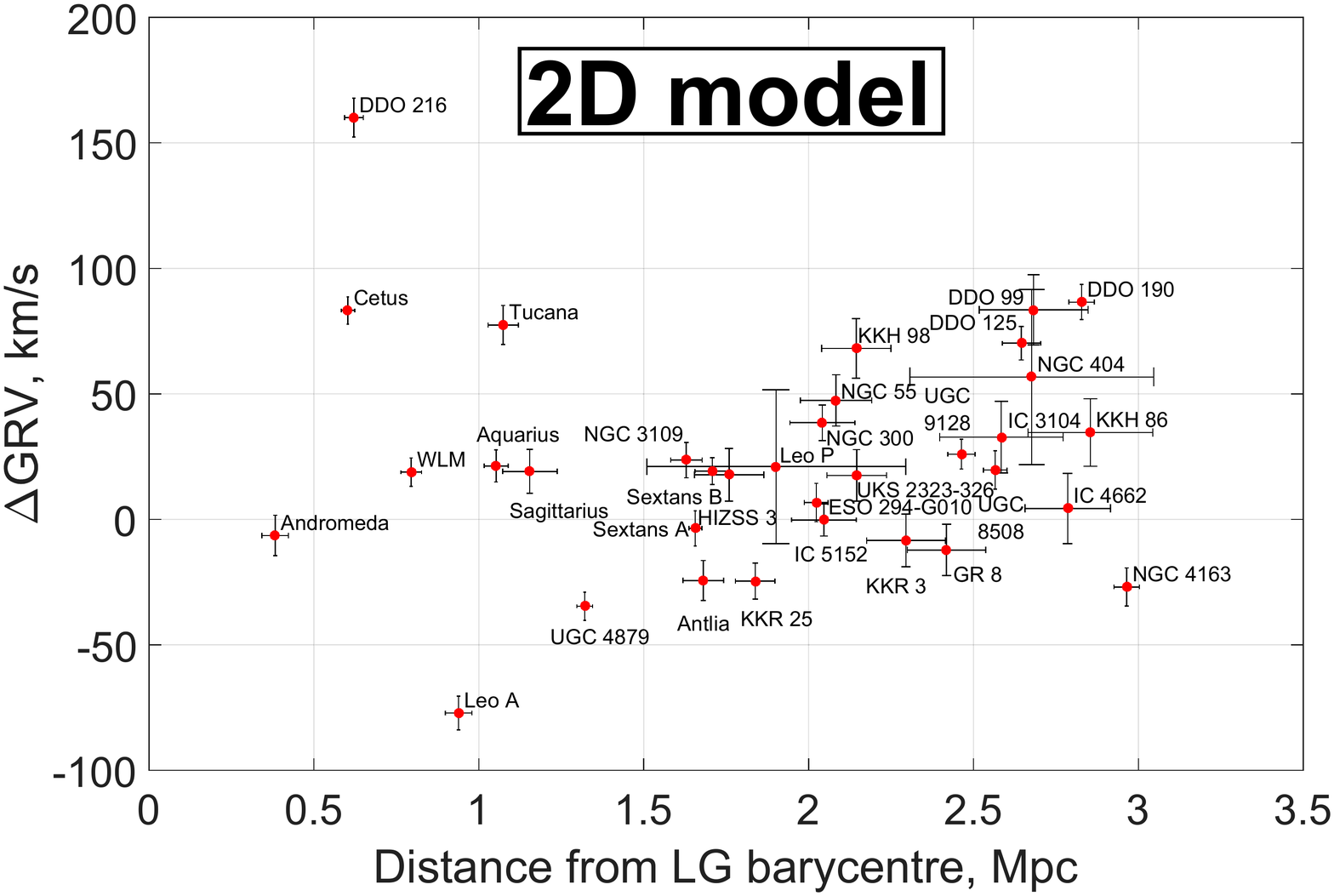}
		\includegraphics [width = 8.5cm] {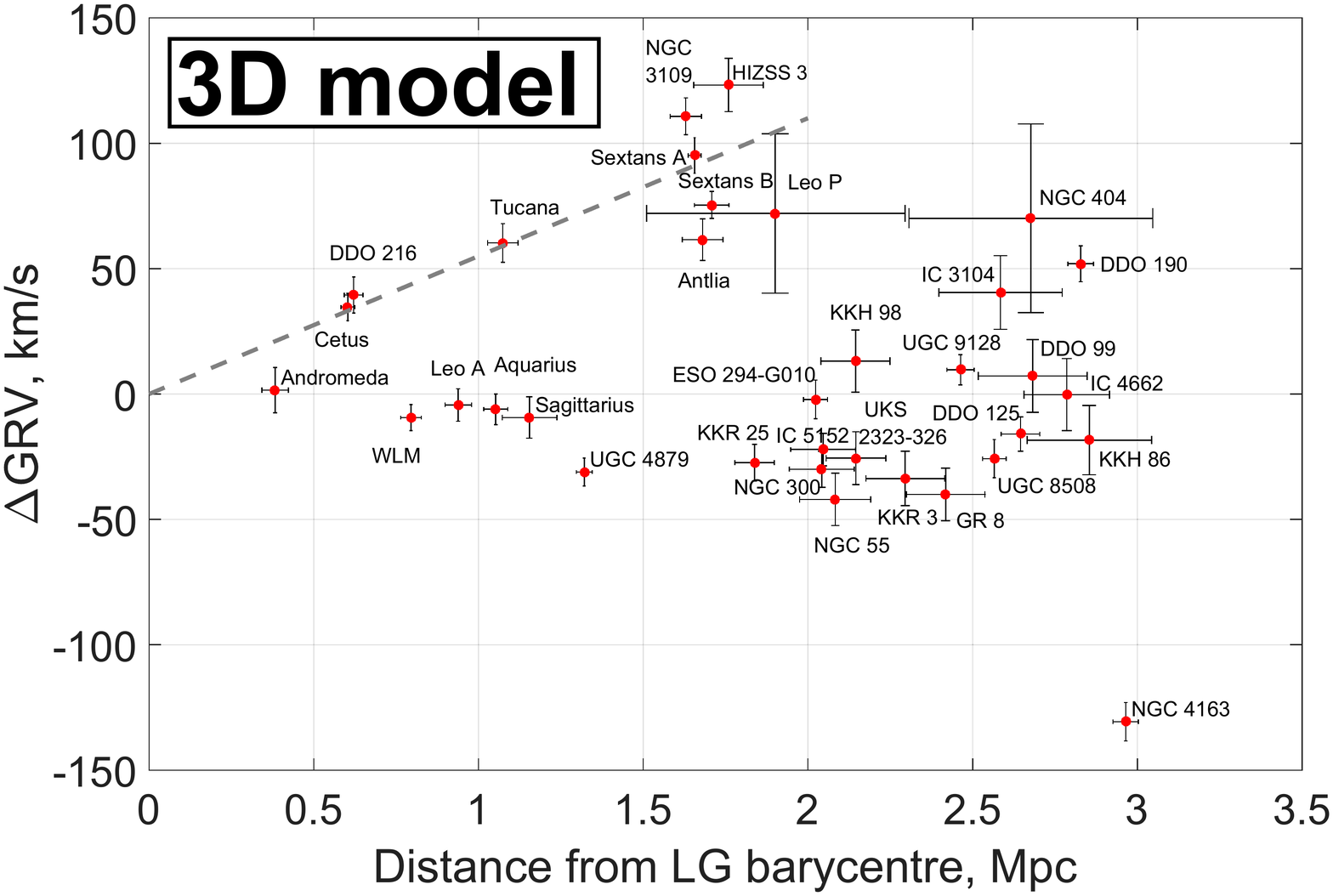}
	\caption{$\Delta GRV$ is shown for our target galaxies against their distance from the LG barycentre. Parameters of the models used are given in Table \ref{Best_fit_parameters}, with the best-fitting ones used for the relevant number of dimensions in the model. Errors shown tend to be anti-correlated because a larger distance to a target increases its predicted GRV, reducing its $\Delta GRV$.}
	\label{Distance_GRV_correlation}
\end{figure}

For this explanation to work, the galaxies with a high GRV in our $\Lambda$CDM-based model need to have been flung out from close to the LG barycentre at around the time the MW and M31 had their interaction. This implies that galaxies with a higher $\Delta GRV$ should generally lie further away from the LG. Thus, it is interesting that the conclusions we reached above using NGC 3109 also hold with Tucana because its lower $\Delta GRV$ is compensated by a smaller distance from the LG. This can be seen visually if one draws a line through these galaxies on Figure \ref{Distance_GRV_correlation} and realises that it passes close to the origin.

The $\Delta GRV \appropto d$ relation between velocities and distances $d$ from the LG barycentre also seems to apply to HIZSS 3, Sextans A and Sextans B. In theory, it applies to Cetus and DDO 216, though their low values of $\Delta GRV$ may mean that this is just a coincidence. Antlia falls slightly below this relation, but its GRV might have been reduced due to the effects of NGC 3109. Tidal features in it suggest that the two may have already undergone a pericentre \citep{Barnes_2001}. The distance to Leo P is still sufficiently uncertain that it is consistent with the $\Delta GRV \appropto d$ relation.

NGC 404 appears to have just as large a $\Delta GRV$ as some of the galaxies just considered, despite being further from the LG. However, we showed in Section \ref{Great_Attractor} that its GRV has likely been increased by $\ssim 15$ km/s due to the GA (Figure \ref{Great_Attractor_Peebles_Effect_76}) while the VC should not have affected its GRV much (Figure \ref{Virgo_Effect_Peebles_76}). Thus, although the assumed 3.06 Mpc distance to NGC 404 seems to be correct \citep{Dalcanton_2009}, its GRV is not that unusual in a $\Lambda$CDM context.

It is interesting that out of the 18 targets we have at distances of 2$-$3 Mpc from the LG barycentre, all of them are broadly consistent with expectations based on $\Lambda$CDM because none of them have a $1\sigma$ lower bound on their $\Delta GRV$ exceeding 60 km/s.\footnote{NGC 4163 has $\Delta GRV < -60$ km/s, but we focus here on the frequency of galaxies with $\Delta GRV \gg 0$ as the main problem with our models is the existence of several such objects, whereas there is only one object like NGC 4163 and it is anomalous on model-independent grounds \citep[][Table 2]{Makarov_2013}.} However, at least 4 galaxies like this exist at distances of 1$-$2 Mpc, despite only having 11 targets in this range. The hypothesis that such high velocity galaxies are equally likely to exist in both distance bins can be ruled out at the 3$\sigma$ confidence level. This suggests that the mechanism missing from our models can only cause high $\Delta GRV$s out to $\ssim 2$ Mpc, quite unlike the effect of tides (which should be even stronger at greater distances). Consequently, we favour an explanation inside the LG and suggest that the crucial ingredient missing from our models is a past MW-M31 flyby.








\section{Conclusions}
\label{Conclusions}

We construct axisymmetric and three-dimensional dynamical models of the Local Group (LG) in the standard $\Lambda$CDM cosmological model. Neither is able to provide a good match to the observed positions and velocities of galaxies within $\ssim$1$-$3 Mpc of the LG barycentre (Figure \ref{rms_Delta_GRV}). This is despite our 3D model accounting for quite a large number of massive objects both within and outside the LG (Table \ref{Massive_galaxy_list_3D}) with fairly weak prior constraints on their masses (Equation \ref{chi_sq_contribution_M}).

Both analyses reveal several galaxies with radial velocities (RVs) much higher than model predictions. Galaxies with anomalously low RVs are rare (Figures \ref{Delta_GRV_histogram_LMC_2D} and \ref{Delta_GRV_histogram_3D}). Thus, the high velocity galaxy problem within the LG persists when using a 3D model for it. However, the particular galaxies which have anomalously high RVs are different in the 2D and 3D analyses, with Tucana being the only clear example of a galaxy with a high RV compared to the predictions of both models (Figure \ref{Distance_GRV_correlation}).



We consider several possibilities for why there are so many LG galaxies with such high GRVs. Perhaps the most plausible in a $\Lambda$CDM context is tides raised by objects outside the LG (Section \ref{Great_Attractor}). The Great Attractor seems unable to reconcile the kinematics of these galaxies with our model (Figure \ref{Great_Attractor_Peebles_Effect_76}). The Virgo Cluster may help somewhat, though the overall trend is for it to raise the expected GRV of galaxies which already have an anomalously low GRV and vice versa (Figure \ref{Virgo_Effect_Peebles_76}). Thus, we do not consider it likely that tides would help greatly to explain the unusually high GRVs of several LG galaxies. This is especially true when considering that our model has quite a lot of flexibility to adjust the tides raised on the LG by varying the masses of objects outside it (Equation \ref{chi_sq_contribution_M}).

In the framework of $\Lambda$CDM, our axisymmetric and 3D results suggest that the past motions of the MW and M31 are too slow to explain the observed kinematics of LG galaxies. A similar challenge with high velocity objects also exists in some systems far outside the LG. For example, the high relative velocity of the components of the Bullet Cluster \citep{Tucker_1995} is difficult to reconcile with the gravity of their dark matter halos acting over the age of the Universe \citep[e.g.][]{Thompson_Nagamine_2012, Kraljic_2015}.

An explanation for the Bullet Cluster and other similar objects like El Gordo \citep{Molnar_2015, Ng_2015} might well require a modification to our understanding of gravity on large scales. Indeed, cosmological $N$-body simulations in Modified Newtonian Dynamics \citep[MOND,][]{Milgrom_1983} could give rise to much higher pair-wise velocities \citep{Llinares_2009, Diaferio_2011}.

Closer to home, MOND requires that the MW and M31 have undergone a past close flyby \citep{Zhao_2013}. This is a consequence of their much stronger mutual gravitational attraction and the almost radial nature of their orbit \citep{Van_der_Marel_2012}. Their higher relative velocity would likely help to explain observations of other LG galaxies through gravitational slingshot interactions with LG dwarfs. A past encounter between them could also have led to the formation of tidal dwarf galaxies, some of which might have ended up bound to neither and moving away from the LG at high speed.

Another interesting idea is that of dark matter as a superfluid \citep{Berezhiani_2015, Khoury_2016}. In this model, phonons in this superfluid mediate forces between the baryons in a galaxy. This leads to MOND-like behaviour, helping to explain the observed tight correlation between the distribution of baryons in galaxies and their rotation curves \citep{McGaugh_2016}.

However, galaxies still need to be surrounded by large (${\ssim 200}$ kpc) halos of dark matter in the normal phase to account for weak lensing because the phonon-mediated force does not affect photon trajectories. Thus, the substantial galaxy-galaxy weak lensing signal \citep[e.g.][]{Brimioulle_2013, Milgrom_2013} needs to be explained in much the same way as in $\Lambda$CDM. This means that interacting galaxies must experience strong dynamical friction between their dark matter halos. As a result, the MW and M31 could never have approached closely in the context of this model because they would subsequently merge.

We suggest that such an interaction nonetheless occurred and would help to resolve the high velocity galaxy problem in the LG. More work will be required to test this scenario.

\section{Acknowledgements}

The authors wish to thank P.J.E. Peebles for providing the algorithm used in this contribution. They also thank the referee and Stacy McGaugh for useful comments. IB is supported by a Science and Technology Facilities Council studentship. Results from the \textsc{fortran}-based simulations were analysed and plotted using \textsc{matlab}$^\text{\textregistered}$.


\newpage
\bibliographystyle{mnras}
\bibliography{LGP_bbl}
\bsp
\label{lastpage}
\end{document}